\documentclass[%
 reprint,
 amsmath,amssymb,
 aps,
pra,
noeprint,
superscriptaddress,
]{revtex4-2}

\usepackage[utf8]{inputenc}
\usepackage[T1]{fontenc}

\usepackage{graphicx}
\usepackage{floatrow}  
\makeatletter
\let\caption@@@make@ORI\caption@@@make
\def\caption@@@make{%
  \caption@ifundefined\caption@lfmt{\let\caption@lfmt\caption@labelformat}{}%
  \caption@ifundefined\caption@fmt\caption@format\relax
\caption@@@make@ORI}
\makeatother

\usepackage[caption=false]{subfig}  
\usepackage{hyperref}
\usepackage{color}
\usepackage{natbib}
\usepackage{amsmath}
\usepackage{physics}
\usepackage[normalem]{ulem}
\usepackage[ruled,vlined]{algorithm2e}
\usepackage{enumitem}  

\usepackage{booktabs}
\usepackage{array}
\usepackage{multirow}
\newcolumntype{L}[1]{>{\raggedright\let\newline\\\arraybackslash\hspace{0pt}}m{#1}}
\newcolumntype{C}[1]{>{\centering\let\newline\\\arraybackslash\hspace{0pt}}m{#1}}
\newcolumntype{R}[1]{>{\raggedleft\let\newline\\\arraybackslash\hspace{0pt}}m{#1}}

\usepackage{tikz}
\usepackage{tikz-qtree}
\usetikzlibrary{decorations.pathreplacing}
\usetikzlibrary{fadings}

\usepackage{cleveref}
\crefname{table}{Table}{Tables}
\crefname{figure}{Figure}{Figures}
\crefname{equation}{Eq.}{Equations}
\Crefname{equation}{Equation}{Equations}
\crefname{section}{Section}{Sections}
\crefname{enumi}{Step}{Steps}

\newcommand{\imagi}{\textnormal{i}}

\makeatletter
\def\l@subsubsection#1#2{}
\makeatother

\begin{document}

\title{Non-perturbative analytical diagonalization of Hamiltonians with application to coupling suppression and enhancement in cQED}

\author{Boxi Li}
\email{b.li@fz-juelich.de}
\affiliation{Forschungszentrum Jülich, Institute of Quantum Control (PGI-8), D-52425 Jülich, Germany}
\affiliation{Institute for Theoretical Physics, University of Cologne, D-50937 Cologne, Germany}
\author{Tommaso Calarco}
\affiliation{Forschungszentrum Jülich, Institute of Quantum Control (PGI-8), D-52425 Jülich, Germany}
\affiliation{Institute for Theoretical Physics, University of Cologne, D-50937 Cologne, Germany}
\author{Felix Motzoi}
\email{f.motzoi@fz-juelich.de}
\affiliation{Forschungszentrum Jülich, Institute of Quantum Control (PGI-8), D-52425 Jülich, Germany}

\begin{abstract}
Deriving effective Hamiltonian models plays an essential role in quantum theory, with particular emphasis in recent years on control and engineering problems.
In this work, we present two symbolic methods for computing effective Hamiltonian models: the Non-perturbative Analytical Diagonalization (NPAD) and the Recursive Schrieffer-Wolff Transformation (RSWT). 
NPAD makes use of the Jacobi iteration and works without the assumptions of perturbation theory while retaining convergence, allowing to treat a very wide range of models.
In the perturbation regime, it reduces to RSWT, which takes advantage of an in-built recursive structure where remarkably the number of terms increases only linearly with perturbation order, exponentially decreasing the number of terms compared to the ubiquitous Schrieffer-Wolff method.
In this regime, NPAD further gives an exponential reduction in terms, i.e.~superexponential compared to Schrieffer-Wolff, relevant to high precision expansions.
Both methods consist of algebraic expressions and can be easily automated for symbolic computation.
To demonstrate the application of the methods, we study the ZZ and cross-resonance interactions of superconducting qubits systems.
We investigate both suppressing and engineering the coupling in near-resonant and quasi-dispersive regimes.
With the proposed methods, the coupling strength in the effective Hamiltonians can be estimated with high precision comparable to numerical results.

\end{abstract}

\maketitle

\setcounter{tocdepth}{3}
\tableofcontents

\section{Introduction}

\begin{figure*}[t]
\centering
\includegraphics[width=\textwidth]{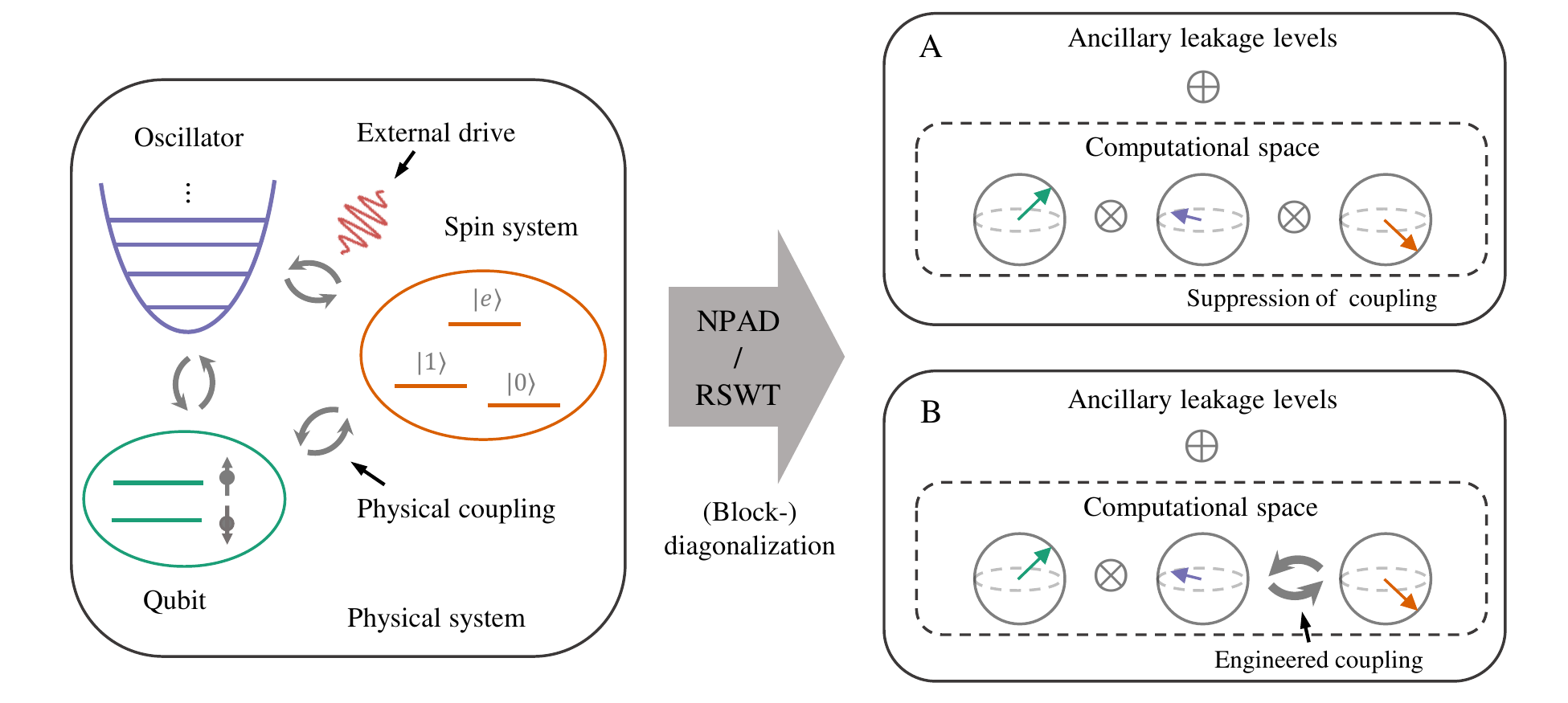}
\caption{Illustration of generating an effective Hamiltonian model from a given physical model. The left-hand side shows the physical system composed of several different quantum subsystems and possible coupling among them. External controls may also exist and drive the system dynamics.
The methods introduced in this article (NPAD and RSWT) can be used to compute the effective model (right-hand side) where undesired interactions are effectively removed (block A) and engineered couplings are enhanced (block B). The dynamics can then be studied in the computational subspace.
}
\label{fig:effective hamiltonian}
\end{figure*}

Deriving effective models is of fundamental importance in the study of complex quantum systems. Often, in an effective model, one decouples the system of interest from the ancillary space, as shown in \cref{fig:effective hamiltonian}. The dynamics are then studied within the effective subspace, which is usually much easier than in the original Hilbert space, and provides fundamental information such as conserved symmetries, entanglement formation, orbital hybridization, computational eigenstates, spectroscopic transitions, effective lattice models, etc. In terms of the Hamiltonian operator, an effective compression of the Hilbert space can be achieved by diagonalization or block diagonalization.

When the coupling between the system and ancillary space is small compared to the dynamics within the subspace, the effective model is often derived by a perturbative expansion. In the field of quantum mechanics, a ubiquitous expansion method that enables reduced state space dimension is the Schrieffer-Wolff Transformation (SWT)~\cite{Schrieffer1966,Bravyi2011}, also known in various sub-fields as adiabatic elimination~\cite{Brion2007}, Thomas-Fermi or Born-Oppenheimer approximation~\cite{Lieb1977,born1927}, and quasi-degenerate perturbation theory~\cite{Suzuki1984}. Finding uses throughout quantum physics, SWT can be found in atomic physics~\cite{Brion2007}, superconducting qubits~\cite{Magesan2020,Malekakhlagh2020}, condensed matter~\cite{Bravyi2011}, semiconductor physics~\cite{Romhanyi2015},  to name a few.

The SWT method is however limited to regimes where a clear energy hierarchy can be found and therefore fails to converge for a wide variety of physical examples. In particular, for infinite-dimensional systems such as coupled harmonic and anharmonic systems (e.g., in superconducting quantum processors), the abundance of both engineered and spurious resonances motivates the use of other techniques. Moreover, even when perturbation theory is applicable, the number of terms in the expansions grows exponentially as the perturbation level and therefore are not practically usable in many instances.

In this article, we introduce a new symbolic algorithm, Non-Perturbative Analytical Diagonalization (NPAD), that allows the computation of closed-from, parametric effective Hamiltonians in a finite-dimensional Hilbert space with a guarantee for convergence. The method makes use of the Jacobi iteration and recursively applies Givens rotations to remove all unwanted couplings. In the perturbative limit, it reduces via BCH expansion to a variant of SWT, which we refer to as the Recursive Schrieffer-Wolff Transformation (RSWT). For this method, the number of commutators grows only linearly with respect to the perturbation order, in contrast to the exponential growth in the traditional approach. Both methods can be used in low-order expansions to provide compact analytical expressions of effective Hamiltonians; or, alternatively, higher-order expansions that allow for fast parametric design~\cite{Goerz2017} and tuning~\cite{Menke2021} of effective Hamiltonian models (and, e.g., subsequent automatic differentiation). As illustrated in \cref{fig:effective hamiltonian}, with the two methods, one can tune the system for engineered decoupling or enhanced controlled coupling.

The key insight of our work is that the iteration step in forming the effective model can be applied recursively, i.e. after each step the transformed Hamiltonian is viewed as a new starting point and determines the next step. Moreover, each step can act on a chosen single state-to-state coupling at a time, thereby providing an exact elimination of the term. In this regard, this can be understood as a generalization of the well-known numerical Jacobi iteration used for diagonalization of real symmetric matrices~\cite{jacobi1846}, which has also found use for Hermitian operators ~\cite{Forsythe1960,henrici1958}.
Similar ideas have also been widely used in the orbital localization problem~\cite{Edmiston1963}.

As demonstrations of the practical utility of the methods, we study superconducting qubits, which are especially relevant for robust parametric design methods, not only because they are prone to spurious resonances~\cite{Malekakhlagh2021,Sank2016,Baker2018}, but because they can be readily fabricated across a very wide range of energy scales \cite{krause2021magnetic, forn2010observation}.

We investigate both the near-resonant regime and in the quasi-dispersive regime, focusing on the ZZ and cross-resonance interaction.
In the near-resonant regime,
we consider the two-excitation manifold and compute accurate approximations of the ZZ interaction strength applicable to the full parameter regime for gate implementation~\cite{Dicarlo2009,Chen2014,Barends2014,Rol2019}.
In the second scenario, we study the suppression of ZZ interactions~\cite{Goerz2017,Zhao2020,Ku2020,Xu2020,Sete2021,Xu2020zz,Zhao2020a,Stehlik2021,Sung2020,kandala2021,Mundada2019,Collodo2020,Chu2021,Jin2021,Finck2021,Wei2021,Mitchell2021,Xiong2021} in the traditional setup of resonator mediated coupling without direct qubit-qubit interaction.
The result shows that the ZZ interaction can be suppressed without resorting to additional coupling in a regime where the qubit-resonator detuning is comparable to the qubit anharmonicity, described by an equation of a circle.
Extending the applications to block diagonalization, we then compute the coupling strength of a microwave-activated cross-resonant interaction.
We show that, with only 4 Givens rotations, we can diagonalize the drive and  achieve accurate estimation in the regime where the perturbation method fails.

This paper is organized as follows:
In \cref{sec:methods}, we present the mathematical methods, NPAD and RSWT, for diagonalization and obtaining effective Hamiltonian models.
We also briefly discuss generalizing the two methods to block diagonalization in \cref{sec:block diagonalization}.
Next, in \cref{sec:applications}, we demonstrate the applications to superconducting systems.
We study the ZZ interaction for generating entanglement in the near-resonant regime (\cref{sec:application1}), and in the (quasi-) dispersive regime for suppressing cross-talk noise (\cref{sec:application2}).
The computation of the cross-resonance coupling strength is presented in \cref{sec:cross resonance}.
We conclude and give an outlook of other possible applications in \cref{sec:conclusion}.

\section{Mathematical methods}
\label{sec:methods}
\subsection{Non-perturbative Analytical Diagonalization}

\label{sec:jacobi}
In this subsection, we introduce the NPAD for symbolic diagonalization of Hermitian matrices and discuss how it can be applied to obtain effective models.

In this algorithm, a Givens rotation is defined in each iteration to remove one specifically targeted off-diagonal term.
By iteratively applying the rotations, the transformed matrix converges to the diagonal form.
The rotation keeps the energy structure when the off-diagonal coupling is small while always exactly removing the coupling even when it is comparable to or larger than the energy gap.
Compared to the Jacobi method used in numerical diagonalization~\cite{jacobi1846, henrici1958, Forsythe1960}, we truncate the iteration at a much earlier stage.
As each iteration consists only of a few algebraic expressions, the algorithm produces a closed-form, parametric expression of the transformed matrix.

We start from a two-by-two Hermitian matrix and define a complex Givens rotation that diagonalizes it.
Then, we generalize the rotation to higher-dimensional matrices, discuss the convergence of the iteration, and how to use it as a symbolic algorithm.
In \cref{sec:application1}, we show a concrete application where we apply NPAD with only two rotations to approximate the energy spectrum of a near-resonant quantum system which can not be studied perturbatively.

\subsubsection{Givens rotations}
\begin{figure}
    \centering
    \subfloat[\label{fig:givens rotation}]{
        \includegraphics[width=0.45\linewidth]{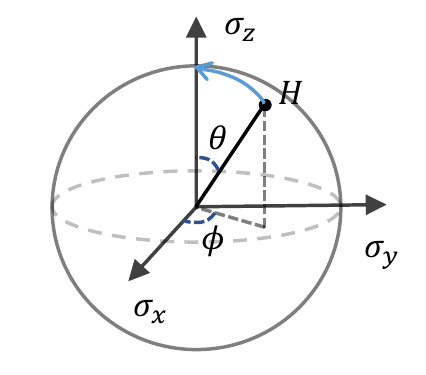}
    }
    \subfloat[\label{fig:computation U}]{
        \includegraphics{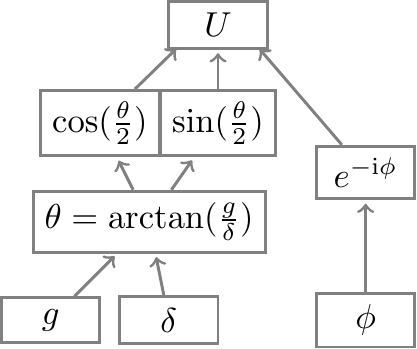}
    }
    \caption{{\bf (a):} The Givens rotation illustrated on a Bloch sphere. A Hermitian matrix defined in \cref{eq:two-by-two hamiltonian} is denoted as a point on the surface of a Bloch sphere with the radius $\sqrt{\delta^2+g^2}$. This is different from the Bloch sphere representation of a quantum state, where the radius is always smaller than or equal to 1. The coordinates correspond to the coefficients in the representation in the Pauli basis.
    The Givens rotation $U$ that diagonalizes the matrix can be viewed as a rotation denoted by the blue arrow (for $\delta \ge 0$).
    {\bf (b):} The computational graph of the Givens rotation $U$, defining the main mathematical steps in the symbolic algorithm~\ref{alg:jacobi}. The inputs $g$, $\delta$ and $\phi$ can be directly extracted from the Hamiltonian. 
    }
\end{figure}

We consider a two-by-two Hermitian matrix
\begin{equation}
    \label{eq:two-by-two hamiltonian}
    H=
    \begin{pmatrix}
    \varepsilon+\delta & g e^{-\imagi\phi} \\
    g e^{\imagi\phi} & \varepsilon-\delta
    \end{pmatrix}
    ,
\end{equation}
where $g$, $\phi$, $\varepsilon$ and $\delta$ are real numbers.
The matrix can be decomposed in the Pauli basis as
\begin{equation}
    H = \epsilon I + \delta \sigma_z + g \left( \cos(\phi) \sigma_x + \sin(\phi) \sigma_y \right)
\end{equation}
which can be illustrated in a Bloch sphere with the radius $\sqrt{\delta^2+g^2}$ (omitting the identity) as shown in \cref{fig:givens rotation}.
Without loss of generality, we assume that $g \geq 0$ and absorb the sign into the complex phase.

The diagonalization can be understood as a rotation on the Bloch sphere to the North or South pole.
In particular, if $\delta \ge 0$, it is rotated to the North pole, and otherwise to the South pole, avoiding unnecessarily flipping the energy level during the diagonalization. 
This rotation is performed around the axis $\hat{n}=\cos(\phi) \sigma_y - \sin(\phi) \sigma_x$  with the angle $\theta=\arctan{(\frac{g}{\delta})}$.
As an illustration, for $\delta \ge 0$, the rotation is denoted by a blue arrow in \cref{fig:givens rotation}.

The unitary transformation that diagonalizes the matrix is given by
\begin{equation}
U = \exp[S] = \exp[\frac{\imagi}{2}\theta \hat{n}]=
\begin{pmatrix}
\cos(\frac{\theta}{2}) & e^{-\imagi\phi}\sin(\frac{\theta}{2}) \\
-e^{\imagi\phi}\sin(\frac{\theta}{2}) & \cos(\frac{\theta}{2}) 
\end{pmatrix}
,
\label{eq:given rotation}
\end{equation}
where $S=\frac{\imagi}{2}\theta  \hat{n}$ is referred to as the generator of the rotation.
The transformation satisfies $\Lambda=U H U^{\dagger}$ with $\Lambda$ the diagonalized matrix.
We refer to $U$ as a Givens rotation~\cite{van1996matrix}.
Notice that in most literature, the Givens rotation is defined with $\phi=0$.
Here we use this more general (Hermitian) definition as it shares many common properties.

The computation of the unitary consists only of elementary mathematical functions, as illustrated in \cref{fig:computation U}.
This is critical for it to be used as a building block for a symbolic algorithm.
As we will see later, by concatenating this building block, a parameterized expression can be generated for an arbitrary Hermitian matrix.

\subsubsection{Simplified formulation}
In practice, the inverse trigonometric function in the expression of $\theta$ is often avoided by using the trigonometric identities
\begin{equation}
\label{eq:trig identity}
    \tan(\theta) = \frac{2t}{1-t^2}
\end{equation}
with $t=\tan(\frac{\theta}{2})$.
We then rewrite \cref{eq:trig identity} as
\begin{equation}
    t^2 + 2t/\kappa - 1 = 0
\end{equation}
with $\kappa=g/\delta$.
We choose the root with smaller norm for the convenience that the rotation will not flip the two energy levels~\footnote{In numerical implementation, it is often written as $t = \frac{\textnormal{sgn}(\kappa)}{|1/\kappa|+\sqrt{1/\kappa^2+1}}$ for numerical stability when $\kappa \rightarrow 0$.}:
\begin{equation}
    \label{eq:jacobi compute t}
    t = \frac{\sqrt{ \kappa ^ 2+1} - 1}{\kappa}
    .
\end{equation}
In this way, the parameters $\cos(\frac{\theta}{2})$ and $\sin(\frac{\theta}{2})$ in the Givens rotation can be calculated directly from $g$ and $\delta$ using algebraic expressions.
It is also evident in \cref{eq:jacobi compute t} that the rotation angle is bounded by $|\theta| \leq \pi/2$.

\subsubsection{The iterative method}
\label{sec:jacobi iteration detail}

We now apply the Givens rotation to remove the ($j,k$)-th entry of a general Hermitian matrix $H$.
The parameters are chosen to be consistent with \cref{eq:two-by-two hamiltonian}, i.e., $\delta_{jk}=(H_{j,j}-H_{k,k})/2$ and $g_{jk}e^{-\imagi\phi_{jk}}=H_{j,k}$.
For simplicity, we use the notation $c_{jk}=\cos(\frac{\theta_{jk}}{2})$, $s_{jk}=\sin(\frac{\theta_{jk}}{2})$, and $t_{jk} = s_{jk}/c_{jk}$.
We write the Givens rotation $U_{jk}$ as
\begin{align}
    U_{jk} = 
    \begin{pmatrix}
     1 &      &      &      &      &      &    \\
      &\ddots&      &      &      &      &   \\
      &      & c_{jk} &\cdots   &e^{-\imagi\phi_{jk}}s_{jk}   &      &   \\
      &      &\vdots&\ddots&\vdots&      &    \\
      &      & -e^{\imagi\phi_{jk}} s_{jk} &\cdots& c_{jk}    &      &   \\
      &      &      &      &      &\ddots&  \\
      &      &      &      &      &      &  1
    \end{pmatrix}
  \label{eq:givens rotation full}
\end{align}
where the diagonal elements are all 1 except for two entries $(j,j)$ and $(k,k)$. All other entries not explicitly defined are 0.

Applying this unitary transformation with $H'=U_{jk}HU_{jk}^{\dagger}$ eliminates the off-diagonal entry $H_{j,k}$, i.e., $|H'_{j,k}| = |H'_{k,j}| = g'_{jk}= 0$. It renormalizes the energies such that
\begin{alignat}{4}
    \label{eq:jacobi elementwise1}
    &\delta'_{jk} &&= \delta_{jk} &&+ t_{jk} g_{jk} 
\end{alignat}
However, this will also mix other entries on the $j,k$-th rows and columns, given by
\begin{alignat}{4}
    \label{eq:jacobi off diagonal1}
    &H'_{h,j} &&= c_{jk} H_{h,j} &&+ e^{\imagi\phi_{jk}}s_{jk}  H_{h,k} \\
    \label{eq:jacobi off diagonal2}
    &H'_{h,k} &&=c_{jk} H_{h,k} &&- e^{-\imagi\phi_{jk}}s_{jk} H_{h,j}
\end{alignat}
with
$h\neq j,k$.

One can diagonalize the matrix by applying the rotation $U_{jk}$ with the corresponding parameters iteratively on the largest remaining non-zero off-diagonal entry, which is referred to as the Jacobi iteration~\cite{jacobi1846}.
That is, we can iteratively solve for the eigenenergies by picking the next largest off-diagonal element, e.g., $H'_{j',k'}=g'_{j'k'}e^{-\imagi\phi'_{j'k'}}$, and applying another Givens rotation, as summarized in \cref{alg:jacobi}.

\begin{algorithm}
\SetAlgoLined
\SetKwInOut{Input}{input}
\SetKwInOut{Output}{output}
\Input{a Hermitian matrix $H_0$}
\Output{an effective model $H'$}
$H \leftarrow H_0$;\\
\While{$\norm{H - \textnormal{diag}(H)}>$ \textnormal{tolerance}}{
    \begin{enumerate}
    \item find the target coupling $H_{j,k}$;
    \item define $\delta_{jk}$, $g_{jk}$ and $\phi_{jk}$ such that \\ $\delta_{jk}=(H_{j,j}-H_{k,k})/2$ and 
    $g_{jk}e^{-\imagi\phi_{jk}}=H_{j,k}$;
    \item $\theta_{jk}\leftarrow\arctan{(\frac{g_{jk}}{\delta_{jk}})}$;
    \item $c_{jk} \leftarrow \cos(\frac{\theta_{jk}}{2})$, $s_{jk} \leftarrow \sin(\frac{\theta_{jk}}{2})$;
    \item define $U$ according to \cref{eq:givens rotation full};
    \item $H \leftarrow UHU^{\dagger}$;
    \end{enumerate}
    }
$H' \leftarrow H$
\caption{Non-Perturbative Analytical Diagonalization (NPAD)}
\label{alg:jacobi}
\end{algorithm}

In practice, the above definition of the Jacobi iteration can be relaxed.
For instance, the next target does not always have to be the largest element. In fact, the order of the rotations does not affect the convergence,  as long as all terms are covered in the iteration rules (e.g., cyclic iterations on all off-diagonal entries)~\cite{Forsythe1960}.
However, performing the rotation first on large elements usually increases the convergence rate.
This can be seen by studying the norm of all off-diagonal terms $\norm{H}_F = \sum_{m \neq n} |H_{m,n}|^2$.
Since we have $|H'_{h,j}|^2+|H'_{h,k}|^2 = |H_{h,j}|^2 + |H_{h,k}|^2$ for $h\neq j, k$ and $H'_{j,k} = 0$,
each Givens rotation reduces the norm of all off-diagonal terms:
\begin{equation}
    \norm{H'}_F = \norm{H}_F - 2 |H_{j,k}|^2
    .
    \label{eq:decreasing norm}
\end{equation}
If ($j,k$) is chosen so that $|H_{j,k}|^2$ is larger than the average norm among the off-diagonal terms, one obtains~\cite{Forsythe1960}
\begin{equation}
    \norm{H'}_F = (1-\frac{2}{N(N-1)}) \norm{H}_F
\end{equation}
where $N(N-1)$ is the total number of off-diagonal terms.
Therefore, the algorithm converges exponentially.
Moreover, if the off-diagonal terms are much smaller than the energy gap, the convergence becomes even faster, i.e., exponentially fast with a quadratic convergence rate~\cite{henrici1958}.
This leads to an efficient variant of the Schrieffer-Wolff-like methods, as described in \cref{sec:RSWT}.

From the above analysis, we also see that the Givens rotation does not have to exactly zero the target coupling.
Instead, it only needs to reduce the total norm.
Therefore, if the structure of the Hamiltonian is known, rotations can be grouped such that all rotations within one group are constructed from the same Hamiltonian and then applied recursively.
We will also explore this possibility in concrete examples later in the article.

As a machine-precision, numerical diagonalization algorithm, the Jacobi iteration is slower than the QR method for dense matrices.
However, in many problems in quantum engineering, the Hamiltonian is often sparse and it is known in advance which interaction needs to be removed.
It is not always necessary to compute the fully diagonalized matrix but only to transform it into a frame where the target subspace is sufficiently decoupled from the leakage levels.
Therefore, an iterative method where each step is targeted at one off-diagonal entry is of particular interest.

As a symbolic method, we can truncate the Jacobi iteration at a very early stage to obtain closed-formed parametric expressions.
It will also correctly calculate the renormalized energy and other couplings while keeping the energy structure in the perturbative limit, as will be discussed in \cref{sec:RSWT}.

\subsection{Recursive Schrieffer-Wolff perturbation method}
\label{sec:RSWT}
In the previous subsection, we introduced NPAD that produces a closed-form, parametric expression of an approximately diagonalized matrix.
Here, we show that in the perturbative limit, where the coupling is much smaller than the bare energy difference, the Jacobi iteration reduces to a Schrieffer-Wolff-like transformation.
Interestingly, the recursive nature of the Jacobi iteration is preserved in this limit.
Instead of looking for one generator that diagonalizes the full matrix as in the traditional Schrieffer-Wolff transformation (SWT), an iteration is constructed such that every time only the leading-order coupling is removed.
We refer to it as recursive Schrieffer-Wolff transformation (RSWT) because of the recursive expression it produces.
We also show that RSWT demonstrates an exponential improvement in complexity compared to SWT for perturbation beyond the leading order.
In \cref{sec:application2}, we demonstrate an application of RSWT in estimating the ZZ interaction between two Transmon qubits in a dispersive regime.

\subsubsection{Givens rotation in the perturbative limit}
In the perturbative limit, compared to $U_{jk}$ in \cref{eq:givens rotation full}, it is more convenient to specify the generator defined in \cref{eq:given rotation}.
For the Givens rotation $U_{jk}$ the corresponding generator $S'$ has two non-zero entries
\begin{align}
    \label{eq:compute generator}
    S'_{j,k} = - {S'_{k,j}}^* = H_{j,k}/(H_{j,j} - H_{k,k})
    ,
\end{align}
all other entries being 0.
In addition,
assuming that we only aim at removing the leading-order off-diagonal terms,
we define a generator 
\begin{equation}
    \label{eq:RSWT generator definition}
    S = \sum_{p\in\mathcal{P}} S'_{p}
\end{equation}
where the sum over $\mathcal{P}$ denotes all pairs of non-zero off-diagonal entries in $H$.
The assumption of perturbation indicates that $\norm{S}_F \ll 1$.
In this case, the unitary generated by $S$ still eliminates all the leading-order coupling because
\begin{equation}
    \label{eq:exp expansion}
    \exp(S) = \exp(\sum_{p \in \mathcal{P}} S'_{p}) = \prod_{p \in \mathcal{P}} e^{S'_{p}} + \mathcal{O}(\norm{S}_F^2)
    .
\end{equation}

This generator $S$ is identical to the generator of the leading-order SWT.
One can verify that $[S, D] = -V$ where $D$ and $V$ are the diagonal and off-diagonal parts of $H$.
By expanding the transformation $e^S H  e^{-S}$ using the BCH formula
\begin{equation}
    H' = e^S H  e^{-S} = H + [S, H] + \frac{1}{2!}[S,[S, H]] + \cdots
    \label{eq:bch formula}
\end{equation}
and truncating the series at $\mathcal{O}(\norm{S}_F^2)$, one obtains the leading-order SWT.

The difference between RSWT and SWT appears when one considers higher-order perturbation.
In SWT, one expands the transformed Hamiltonian $H'$ and the generator $S$ perturbatively as a function of a small parameter and collects terms of the same order on both sides of \cref{eq:bch formula}.
However, here, the generator is predefined and it only eliminates the leading-order coupling.
Similar to the Jacobi iteration, we treat the transformed Hamiltonian $H'$ as a new Hermitian matrix and perform another round of leading-order transformation as the next iteration.
This results in a recursive expression for $H'$, which is still a closed-form expression.
The remaining off-diagonal terms can always be removed by the next iteration if the truncation level of BCH formula is high enough to guarantee sufficient accuracy.
We present the iteration of RSWT in detail in the next subsection and show that it simplifies the calculation for perturbation beyond the leading order.

\subsubsection{The RSWT iterations}
\label{sec:RSWT steps}

In the following, we outline the iterative procedure of the RSWT.
We denote the initial matrix $H$ as step zero, with the notation $D_0 = D$, $ V_0 = \lambda V$ and $H_{0} = H = D_0 + V_0$.
The parameter $\lambda$ is the dimensionless small parameter used to track the perturbation order.
Assume that we want to compute the perturbation to the eigenenergy up to the order $\lambda^K$.
We refer to this as the $\lambda^K$-perturbation.
Given the Hamiltonian of iteration $n$, $H_n$, we can compute the next iteration $H_{n+1}$ as follows.

We first define a generator $S_{n+1}$ according to \cref{eq:RSWT generator definition} such that $[S_{n+1},D_n]=-V_n$, where $D_n$ and $V_n$ are the diagonal and the off-diagonal part of $H_n$.
As the energy gap $D_n$ always stays at $\mathcal{O}(\lambda^0)$ under the assumption of small perturbation, $S_{n+1}$ is of the same order as $V_n$.
Notice that $S_{n+1}$ is generated from the perturbed matrix in the previous iteration, $H_n$, in contrast to the unperturbed matrix as in SWT.

Then, the next level of perturbation is computed with
\begin{equation}
    \label{eq:RSWT equation unsimplified}
    H_{n+1} =
    \sum_{t=0}^m
    \frac{1}{t!}
    \mathcal{C}_t(S_{n+1}, D_n)
    +
    \sum_{t=0}^{m-1}
    \frac{1}{t!}
    \mathcal{C}_t(S_{n+1}, V_n)
\end{equation}
where $\mathcal{C}$ is the nested commutator defined by
\begin{equation}
    \mathcal{C}_{t+1}(A, B) = [A, \mathcal{C}_{t}(A, B)]
\end{equation}
and $\mathcal{C}_{0}(A, B) = B$.
The truncation level $m$ of the BCH expansion will be defined explicitly later.
Because $[S_{n+1},D_n]=-V_n$ by construction, we have for all $n$ and $t$
\begin{equation}
    \label{eq:cancellation}
    \mathcal{C}_{t+1}(S_{n+1},D_n)
    =-\mathcal{C}_t(S_{n+1}, V_n)
    .
\end{equation}
Therefore, plugging in \cref{eq:cancellation} into \cref{eq:RSWT equation unsimplified} simplifies it to
\begin{equation}
    \label{eq:RSWT equation}
    H_{n+1} = D_n +
    \sum_{t=1}^{m-1}
    \frac{t}{(t+1)!}
    \mathcal{C}_t(S_{n+1}, V_n)
    .
\end{equation}
Notice that $t$ starts from 1 in the sum, which means that all coupling terms at the same order of $V_n$ are removed and the order of the remaining coupling, $V_{n+1}$, is squared.
This iteration is applied until the desired order is reached, as summarized in \cref{alg:RSWT}.

\begin{algorithm}
\SetAlgoLined
\SetKwInOut{Input}{input}
\SetKwInOut{Output}{output}
\Input{a Hermitian matrix $H_0$}
\Output{$H'$ including correction to the eigenenergy up to $\lambda^K$}
$n_{\rm{max}} \leftarrow \lfloor \log_2(K)\rfloor$\;
\For{$n \leftarrow 0;\ n < n_{\rm{max}};\ n \leftarrow n + 1$}{
    \begin{enumerate}
    \item $D_n \leftarrow \textnormal{diag}(H_{n})$;
    $V_n \leftarrow H_{n} - D_n$;
    \item 
    initialize a zero matrix $S_{n+1}$\;
    \SetAlgoNoLine
    \For{$j, k$ \textnormal{with} $V_{n,j,k} \neq 0$}
    {
        $S_{n+1, j, k} \leftarrow V_{n,j,k}/(D_{n,j,j} - D_{n,k,k})$\;
    }
    \item $m \leftarrow \lfloor\frac{K}{2^{n}} \rfloor$\;
    $H_{n+1} \leftarrow D_n +
        \sum_{t=1}^{m-1}
        \frac{t}{(t+1)!}
        \mathcal{C}_t(S_{n+1}, V_n)$\;
    \end{enumerate}
    }
$H' \leftarrow H_{n_{\textnormal{max}}}$
\caption{Recursive Schrieffer-Wolff Transformation (RSWT)}
\label{alg:RSWT}
\end{algorithm}

To ensure that the truncation of the BCH is accurate up to the order $\mathcal{O}(\lambda^{K})$,
for the $n$th iteration, we need to choose the truncation $m = \lfloor\frac{K}{2^{n}} \rfloor$, which ensures that $H_{n+1} = e^{(S_{n+1})}H_{n}e^{(-S_{n+1})} + \mathcal{O}(\lambda^{K+1})$.
This maximal level $m$ is halved every time the iteration step increases because the remaining coupling is quadratically smaller.
This means that, in contrast to SWT, the first iteration has the largest number of terms in RSWT.
In \cref{sec:RSWT error bound}, we show that, if $\norm{S_{n+1}}<\frac{1}{2}$, the error of the truncation in \cref{eq:RSWT equation} is bounded by
\begin{equation}
    \norm{
        H_{n+1} - H_{n+1}^{\infty}
    }
    \le
    \frac{2^{m}}{m!}
    \norm{S_{n+1}}^m
    \norm{V_{n}}
\end{equation}
where the $H_{n+1}^{\infty}$ is \cref{eq:RSWT equation} in the limit $m\rightarrow \infty$.

\subsection{Block diagonalization}
\label{sec:block diagonalization}
Both the NPAD and the RSWT methods introduced in the previous sections can be designed to only target a selected set of off-diagonal terms and, hence, used for block-diagonalization. This is especially useful to engineer transversal coupling in a subsystem and leave the remaining levels as intact as possible. Here, we briefly discuss these generalizations. Notice that it is always possible to first diagonalize the matrix and then reconstruct the block diagonalized form that satisfies certain conditions, for instance as in Ref.~\cite{Cederbaum1989}. In the following, we discuss only methods that do not diagonalize the matrix first.

In NPAD, by construction, each rotation removes one off-diagonal element.
With Givens rotations only applied to the inter-block elements, an iteration for block diagonalization can be defined.
The norm of all off-diagonal entries, $\norm{H}_F$, is still monotonously decreasing according to \cref{eq:decreasing norm}.
Hence, a limit exists and its convergence is also the convergence of the block diagonalization.
However, the convergence is not always monotonous with respect to the norm of all inter-block terms.
This is because a Givens rotation may rotate a large intra-block term into an inter-block entry.
Therefore, the algorithm may not always converge faster than the full diagonalization would.
Nevertheless, if the dominant coupling terms in the Hamiltonian are known, the Jacobi iteration can be designed to target at those to realize an efficient block-diagonalization.
In \cref{sec:cross resonance}, we show an example of this in computing the cross-resonance coupling strength through NPAD.

For perturbation, RSWT can be applied as a block diagonalization method under the constraint that both the inter-block and the intra-block coupling are much smaller than the inter-block energy gap.
This can be achieved by slightly modifying the RSWT iterations:
We first separate the diagonal, the intra-block and the inter-block terms: $H_n = D_n + V^{\rm{intra}}_n +  V^{\rm{inter}}_n $.
Next, in \cref{alg:RSWT} we only define $S$ for those non-zero entries in $V^{\textnormal{inter}}_n$, i.e. the couplings we wish to remove.
And in the last step we replace \cref{eq:RSWT equation} with 
\begin{align}
    H_{n+1} = 
    D_n
    &+
    \sum_{t=1}^{m-1}
    \frac{t}{(t+1)!}
    \mathcal{C}_t(S_{n+1}, V^{\rm{inter}}_n) \nonumber \\
    &+
    \sum_{t=0}^m
    \frac{1}{t!}
    \mathcal{C}_t(S_{n+1}, V^{\rm{intra}}_n)
    .
\end{align}
In this definition, the leading inter block coupling is of the order $\mathcal{O}([S_{n+1}, V^{\rm{intra}}_n])$.
As we do not remove the intra-block coupling, we still get $V^{\rm{intra}}_n=\mathcal{O}(\lambda)$.
Therefore, the remaining coupling is $\mathcal{O}(\lambda  V^{\rm{intra}}_n)$, i.e. the perturbation order is increased by one, instead of being squared as in the case of full diagonalization.
Therefore, the exponential reduction of the number of commutators does not always apply in the case of block diagonalization.
However, notice that the small parameter $\lambda$ here is defined as the (largest) ratio between the inter-block couplings and gaps, which is usually much smaller than those within the block.
Hence, if carefully designed, the convergence can still surpass the full diagonalization in the first few perturbative orders.

\subsection{Comparison between different methods}
To help understand the proposed methods, we here discuss the difference between them and the traditional methods.
We first compare RSWT with traditional SWT and then NPAD with the perturbation methods.

For RSWT, with the same target accuracy, e.g., $\mathcal{O}(\lambda^{K})$, it should provide the same expression as from SWT, up to the error $\mathcal{O}(\lambda^{K+1})$.
However, compared to the SWT, RSWT requires a much smaller number of iterations and commutators.
To reach $\mathcal{O}(\lambda^K)$, SWT needs $K-1$ iterations, while RSWT only needs $\lfloor\log_2(K) \rfloor$ because of the quadratic convergence rate.
More importantly, the total number of commutators grows only linearly for RSWT, compared to the exponentially fast growth for SWT~\cite{Magesan2020}.

Intuitively, this is because RSWT uses the recursive structure and avoids unnecessary expansions of the intermediate results.
Mathematically, this can be seen from the following two aspects:
First, in RSWT, each iteration improves the perturbation level from $\lambda^k$ to $\lambda^{2k}$, instead of $\lambda^{k+1}$.
Hence, the number of iterations increases only logarithmically with respect to the perturbation order, as seen in the definition of $n_{\rm{max}}$ in \cref{alg:RSWT}.
This is because we always treat the transformed matrix as a new one and remove the leading-order coupling.
It is consistent with the quadratic convergence rate of the Jacobi iteration with small off-diagonal terms.
Second, in RSWT, the generator $S_n$ is only used at the current iteration.
Hence, there are no mixed terms such as $[S_2, [S_1, V_0]]$, in contrast to SWT.

The total number of commutators required to reach level $\lambda^K$ is shown in \cref{tab:number of commutators},
where we have taken into consideration that if $\mathcal{C}_t(A,B)$ is known, computing $\mathcal{C}_{t+1}(A,B)$ only requires one additional commutator.
The detailed calculation is presented in \cref{sec:number of commutators}.

The NPAD method, on the other hand, uses non-linear rotations to replace the linear perturbative expansion.
More concretely, in the Jacobi iteration, by targeting only one coupling in each recursive iteration, the unitary transformation can be analytically expressed as a Givens rotation, thus avoiding the BCH expansion in \cref{eq:bch formula}.
Therefore, it efficiently and accurately captures the non-perturbative interactions in the system.

To compare it with the perturbation methods, we estimate the number of operations required for NPAD in the perturbative regime.
Assume we construct the Jacobi iteration from the $G$ coupling terms used in generating an $S$ in RSWT.
Applying those unitaries is, to the leading order, the same as applying one RSWT iteration.
A single Givens rotation on a Hamiltonian takes $\mathcal{O}(N)$ operations, where $N$ is the matrix size.
Thus, the cost for computing the effective Hamiltonian after $G$ rotations is the same as computing one commutator $[S,\boldsymbol{\cdot}]$, up to a constant factor.
Because the Givens rotation avoids the BCH expansion, there is no nested commutators and the total number operations is $\mathcal{O}(n_{\rm{max}}NG)$  with $n_{\rm{max}}$ the number of iterations in \cref{alg:RSWT}.
Hence the number of terms scales logarithmically with respect to $K$ instead of linearly as for RSWT, i.e., a super-exponential reduction compared to SWT (\cref{tab:number of commutators}).
However, the non-linear expressions provided by NPAD are usually harder to simplify and evaluate by hand compared to the rational expressions obtained from perturbation.

From the above discussion, one can see that it is also straightforward to combine NPAD with perturbation.
Instead of fully diagonalizing the matrix, the Jacobi iteration can be designed to remove only the dominant couplings and combined with perturbation methods to obtain simplified analytical expressions.
In fact, this is often used implicitly in the analysis when, e.g., a strongly coupled two-level system is perturbatively interacting with another quantum system.
The Jacobi iteration suggests that this can be generalized systematically to more complicated scenarios.

{
\setlength\heavyrulewidth{0.35ex}
\begin{table}[h]
\begin{tabular}{ l  C{0.53cm} C{0.53cm} C{0.53cm} C{0.53cm} C{0.53cm}C{0.53cm}C{0.53cm}}

\toprule
$K$ & 2 & 3 & 4 & 5  & 6  & 7  & 8   \\
\midrule
SWT   & 1 & 4 & 11 & 26  & 57  & 120 & 247  \\[0.1cm]
RSWT   & 1 & 2 & 4 & 5  & 7  & 8 & 11  \\[0.1cm]
NPAD  & 1 & 1 & 2  & 2  & 2  & 2 & 3 \\
\bottomrule
\end{tabular}
\caption{
The number of terms in the evaluation for different methods to reach the $\lambda^K$-perturbation.
The number denotes the total number of commutators in SWT and RSWT, or the total number of sweeps over all couplings for NPAD. This describes both the "algebraic complexity" (i.e~complexity of the output algebraic expressions) and the computational (time-cost) complexity.
The complexity is reduced from exponential to linear and eventually to logarithmic.
However, notice that although the computational complexity for one commutator and for one Jacobi sweep scales the same in terms of the number of couplings to be removed (see the main text), the Givens rotation in NPAD consists of non-linear algebraic expressions which are individually more expensive to compute.
\label{tab:number of commutators}
}
\end{table}
}

\section{Physical applications}
\label{sec:applications}
In this section, we use the methods introduced in \cref{sec:methods} to study the ZZ interaction in two different parameter regimes.
In a two-qubit system, the ZZ interaction strength is defined by
\begin{equation}
    \label{eq:ZZ interaction def}
    \zeta = E_{11} - E_{10} - E_{01} + E_{00}  
\end{equation}
where $E_{jk}$ denotes the eigenenergy of the two-qubit states $\ket{jk}$.
The Hamiltonian interaction term is written as $\zeta\sigma_{z_1}\sigma_{z_2}$, acting on the two qubits.
Typically, in superconducting systems, it arises from the interaction of the $\ket{11}$ state with the non-computational state in the physical qubits, and can both be used as a resource for entangling gates~\cite{Dicarlo2009,Chen2014,Barends2014,Rol2019} or viewed as cross-talk noise that needs to be suppressed~\cite{Zhao2020,Ku2020,Xu2020,Sete2021,Xu2020zz,Zhao2020a,Stehlik2021,Sung2020,kandala2021,Mundada2019,Collodo2020,Chu2021,Jin2021,Finck2021,Wei2021,Mitchell2021,Xiong2021}.

\subsection{Effective ZZ entanglement from multi-level non-dispersive interactions}
\label{sec:application1}
In this first application, we apply the NPAD method described in \cref{sec:jacobi} to study a model consisting of two directly coupled qubits in the near-resonant regime, where the ZZ interaction can be used to construct a control-Z (CZ) gate (see \cref{fig:effective hamiltonian} block B)~\cite{Dicarlo2009,Chen2014,Barends2014,Rol2019}.
We show that, with two rotations, NPAD provides an improvement on the estimation of the interaction strength for at least one order of magnitude, compared to approximating the system as only a single avoided crossing between the strongly interacting levels, as is standard in the literature.
In addition, if one of the non-computational bases is comparably further detuned than the other, the correction takes the form of a Kerr nonlinearity, with a renormalized coupling strength accounting for the near-resonant dynamics.

\begin{figure}
    \centering
    \subfloat[\label{fig:CZ illustration}]{
        \includegraphics[width=.23\linewidth]{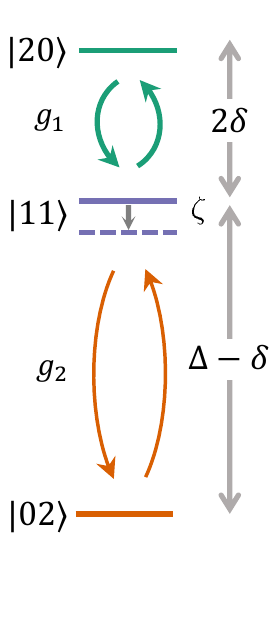}
    }
    \subfloat[\label{fig:zeta with jacobi}]{
    \includegraphics[width=.72\linewidth]{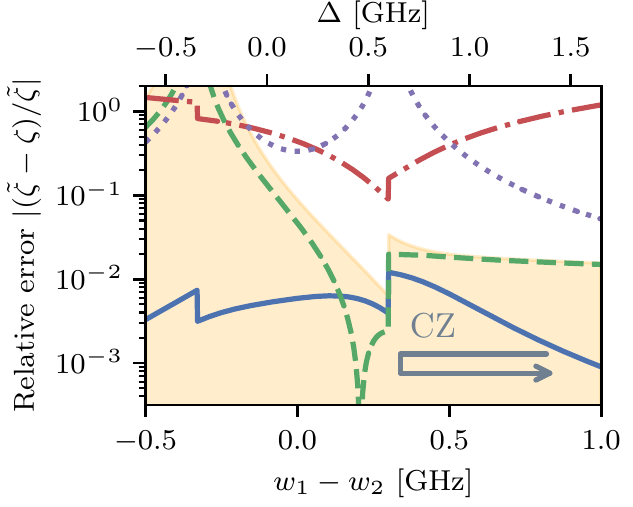}
    }

    \caption{
    {\bf (a):} Interaction and energy level diagram of the two-excitation manifold in the unperturbed Hamiltonian given by \cref{eq:sc qubit-qubit model}.
    The solid lines represent the bare qubit states, while the arrow and the dashed purple line denote the Stark shift and the eigenenergy of the perturbed $\ket{11}$ state.
    {\bf (b):}
    Performance of ZZ interaction estimation using NPAD.
    We plot the relative difference between the estimated $\zeta$ and the value obtained by numerical diagonalization $\tilde{\zeta}$.
    The estimations are computed with 2 rotations (solid, \cref{eq:zeta two rotation}), hybrid method with the additional assumption $\Delta \gg \delta, g_j$ (dashed, \cref{eq:zeta two rotation + approximation}), by assuming only a 2-level system (dash-dot, \cref{eq:zeta 2 level}), and with a leading-order perturbation (dotted).
    The shaded area covers the region below the error estimation given by \cref{eq:jacobi example error}.
    The grey arrow denotes a typical path to generate a CZ gate through ZZ interaction by changing the qubit-qubit detuning.
    The two jumps are located at $\omega_1=\omega_2+\alpha_2$ and $\omega_1+\alpha_1=\omega_2$, i.e., the points where the bare energy level swaps.
    This changes the direction of the Givens rotation.
    The parameters used are $g_1=g_2=\sqrt{2}\cdot 0.1$~GHz and $\alpha_1=\alpha_2=-0.3$~GHz.
    }
\end{figure}

We consider the Hamiltonian of two superconducting qubits that are directly coupled under the rotating-wave \cite{Motzoi2013, zeuch2020exact} and Duffing \cite{khani2009optimal} approximations:
\begin{align}
    H &= \sum_{q \in \{1,2\}}\omega_q b_q^{\dagger} b_q + \frac{\alpha_q}{2} b_q^{\dagger} b_q^{\dagger} b_qb_q + g (b_1 b_2^{\dagger} + b_1^{\dagger} b_2)
\label{eq:sc duffing model}
\end{align}
where $b_q$, $\omega_q$, $\alpha_q$ are the annihilation operator, the qubit bare frequency and the anharmonicity, respectively. The parameter $g$ denotes the coupling strength.
In this Hamiltonian, the sum of the eigenenergies is always a constant $E_{10} + E_{01} = \omega_1 + \omega_2$ because of the symmetry.
Hence, the ZZ interaction comes solely from the interaction between the state $\ket{11}$ and the non-computational basis $\ket{20}$ and $\ket{02}$.
If the frequency is tuned so that the state $\ket{11}$ is close to one of the non-computational states, the coupling will shift the eigenenergy, leading to a large ZZ interaction (\cref{fig:CZ illustration}).

For simplicity, we consider the Hilbert subspace consisting of $\ket{20}$, $\ket{11}$, $\ket{02}$ and write the following Hamiltonian
\begin{equation}
H=
    \begin{pmatrix}
    \delta  & g_1 & 0 \\
  g_1& -\delta & g_2 \\
    0 & g_2 & -\Delta
    \end{pmatrix}
    .
\label{eq:sc qubit-qubit model}
\end{equation}
The parameters in the diagonal elements are given by $\delta = (\omega_1-\omega_2+\alpha_1)/2$ and  $\Delta = 3(\omega_1- \omega_2)/2 - \alpha_2+\alpha_1/2$.
To keep the result general, we use two different coupling strengths $g_1$ and $g_2$, although according to \cref{eq:sc duffing model} they both equal $\sqrt{2}g$.
Without loss of generality, we assume the state $\ket{02}$ is comparably further detuned from the other two, i.e. $\Delta > g_j, \delta$.
If in contrast $\ket{20}$ is further detuned, one can exchange the $\ket{02}$ and $\ket{20}$ in the matrix and redefine $\delta$ and $\Delta$ accordingly.
Notice that this Hamiltonian is different from a $\Lambda$ system~\cite{Brion2007}, where coupling exists only between far detuned levels.

To implement the CZ gate, one tunes the qubit frequency $\omega_1$ so that the states $\ket{11}$ and $\ket{20}$ are swept from a far-detuned to a near-resonant regime.
Hence, the perturbative expansion diverges and cannot be used.
A naive approach is to neglect the far-detuned state $\ket{02}$ and approximate the interaction as a single avoided crossing.
In this case, $\zeta$ is approximated by
\begin{equation}
    \label{eq:zeta 2 level}
    \zeta_{\textnormal{2-level}} \approx \delta - \delta\sqrt{1 + \frac{g_1^2}{\delta^2}}
    .
\end{equation}
However, the interaction $g_2$ results in an error that, in the experimentally studied parameter regimes, can be as large as 10\%, as shown in \cref{fig:zeta with jacobi}.

In the following, we show that with only two Givens rotations, one can obtain an analytical approximation, with the error reduced by one order of magnitude.
The correction can be understood as a Kerr non-linearity with a renormalized coupling strength.

To get an accurate estimation of the ZZ interaction $\zeta$, we need to calculate the eigenenergy of $\ket{11}$ by eliminating its coupling with the other two states.
Therefore, we will make two rotations sequentially on the entry $(0,1)$ and $(1,2)$, given by
\begin{equation}
    H^{(2)} = U_2 H^{(1)} U_2^{\dagger} = U_2 U_1 H U_1^{\dagger} U_2^{\dagger}
    ,
\end{equation}
where $U_1$ and $U_2$ are Givens rotations (\cref{eq:given rotation}) constructed for eliminating the entries $(0,1)$ and $(1,2)$.
Because the matrix is real symmetric, the phase $\phi$ in \cref{eq:given rotation} is 0.

The first transformed Hamiltonian, $H^{(1)} = U_1 H U_1^{\dagger} $, takes the form
\begin{equation}
H^{(1)} = 
\begin{pmatrix}
    E_2& 0& g_2s_{01} \\
    0& -E_2& c_{01}g_2 \\
    g_2s_{01}& c_{01}g_2& -\Delta
\end{pmatrix}
\end{equation}
where $E_2 = \delta\sqrt{1 + \frac{g_1^2}{\delta^2}}$ is the eigenenergy for diagonalizing the two-level system of $\ket{20}$ and $\ket{11}$, consistent with \cref{eq:zeta 2 level}.
The notations used is the same as in \cref{sec:jacobi iteration detail}.
In this frame, the coupling between $\ket{11}$ and $\ket{02}$ is reduced to $c_{01}g_2$, where $c_{01}$ is given by the non-linear expression
\begin{equation}
    c_{01} = 
    \frac{1} {
        \sqrt{
        \left(
        \frac{E_2-\delta}{g_1}
        \right)
        ^2 + 1
        }
    }
    .
\end{equation}
This non-linearity is crucial for the accurate estimation of the eigenenergy.

The second rotation further removes this renormalized coupling $c_{01}g_2$, giving
\begin{equation}
H^{(2)} = 
\begin{pmatrix}
    E_2 & g_2 s_{01} s_{12}& c_{12} g_2 s_{01}\\
    g_2 s_{01} s_{12}& - E_2 + g_2c_{01} t_{12} & 0\\
    c_{12} g_2 s_{01}& 0& -\Delta - g_2c_{01} t_{12} 
\end{pmatrix}
.
\end{equation}
Including the new correction, $g_2c_{01} t_{12}$, the eigenenergy of state $\ket{11}$ reads
\begin{equation}
    \label{eq:zeta two rotation}
    H^{(2)}_{1,1} =
    -E_2 + \frac{\Delta - E_2}{2} \left(\sqrt{1 + \left(\frac{2 c_{01} g_2}{\Delta - E_2} \right)^2}-1 \right)
    .
\end{equation}
In \cref{fig:zeta with jacobi}, we plot the error of the estimated interaction strength $\zeta = H'_{1,1}+\delta$ using typical parameters of superconducting hardware, compared to the numerical diagonalization $\tilde{\zeta}$.
An improvement of at least one order of magnitude is observed compared to traditional methods.

Following the assumptions that $\Delta \gg \delta, g_j$, \cref{eq:zeta two rotation} simplifies to 
\begin{equation}
    \label{eq:zeta two rotation + approximation}
    \boxed{
    H^{(2)}_{1,1} \approx
    - E_2 + \frac{c_{01}^2 g_2^2}{\Delta - E_2}
    }
    .
\end{equation}
We see that the correction takes the form of a Kerr non-linearity~\cite{Holland2015}, but with a renormalized coupling strength $c_{01} g_2$.
This non-linear factor $c_{01}$ accounts for the dynamics between $\ket{20}$ and $\ket{11}$ in the near-resonant regime.
The same effect can be observed in higher levels where similar three-level subspaces exist.
This approximation is plotted as a dashed curve in \cref{fig:zeta with jacobi}.

The error of this estimation comes both from the expansion of the square root in \cref{eq:zeta two rotation} as well as from the remaining coupling in $H^{(2)}$.
The former can be approximated by the next order expansion
\begin{equation}
    \label{eq:jacobi example error}
    \epsilon_1 \approx \frac{c_{01}^4 g_2^4}{(\Delta - E_2)^3}
    .
\end{equation}
For the latter, we consider the remaining coupling in $H^{(2)}$ between $\ket{20}$ and $\ket{11}$, which reads $g_2 s_{01} s_{12}$.
In the limit $\Delta \gg \delta, g_j$, we have
$s_{12} \le
\frac{\theta_{12}}{2} \le
\frac{g_2}{\Delta-E_2} \ll 1$,
indicating that this coupling is much smaller than the energy difference.
Hence, further correction can be estimated by
\begin{equation}
    \epsilon_2 \approx
    \frac{{\left(H^{(2)}_{0,1}\right)}^2}{|H^{(2)}_{0,0} - H^{(2)}_{1,1}|} \le \frac{(g_2 s_{01} s_{12})^2}{g_1} \le \frac{g_2^4  s_{01}^2}{g_1(\Delta-E_2)^2}
    .
\end{equation}
The contribution of the other remaining coupling between $\ket{20}$ and $\ket{02}$ is much smaller due to the large energy gap.
Since $\epsilon_2$ is one order smaller than the $\epsilon_1$, $\epsilon_1$ will be the dominant error.
We plot the region below this error in \cref{fig:zeta with jacobi} as a shaded background.

For the more general cases without assuming $\Delta \gg \delta, g_j$, it is hard to provide an error estimation due to the non-linearity.
However, the result in \cref{fig:zeta with jacobi} indicates that \cref{eq:zeta two rotation} still shows a good performance in other parameter regimes commonly used in superconducting hardware, with an error smaller than 3\%.
We also observe that an improvement for another order of magnitude can be achieved by introducing a third rotation again on the entry $(0,1)$.

\subsection{ZZ coupling suppression in the quasi-dispersive regime}
\label{sec:application2}
In this second example, we use the two methods to investigate the suppression of ZZ cross-talk with the qubit-resonator-qubit setup in the dispersive cQED regime, which corresponds to \cref{fig:effective hamiltonian} block A.
We demonstrate that in the traditional setup without direct inter-qubit coupling, the ZZ interaction defined in \cref{eq:ZZ interaction def} can still be zeroed in a quasi-dispersive regime by engineering the two parameters of qubit-resonator detuning.
The zero points are described by an equation of a circle in the $\lambda^4$-perturbation.
To accurately capture the interaction strength in the quasi-dispersive regime, we also compute with RSWT the $\lambda^6$-perturbation and show that the NPAD method with only 8 Givens rotations provides an expression with similar accuracy.

We consider a Hamiltonian of two superconducting qubits connected by a resonator:
\begin{align}
    \label{eq:qubit-resonator-qubit ham}
    H = & \sum_{q \in \{1,2\}} \omega_q b_q^{\dagger} b_q+ 
    \frac{\alpha_q}{2} b_q^{\dagger} b_q^{\dagger} b_qb_q
    + g_q (b_q a^{\dagger} + b_q^{\dagger} a)\\ \nonumber
    &+ \omega_{\rm{r}} a^{\dagger} a
    .
\end{align}
Due to the finite detuning between the resonator and the qubits, a static ZZ interaction exists even if there is no additional control operation on the system.
In order to implement high-quality quantum operations, this interaction needs to be sufficiently suppressed.

Several approaches have been developed to suppress the ZZ interaction.
One way is to add a direct capacitive coupling channel in parallel with the resonator~\cite{Mundada2019,Xu2020,kandala2021,Sung2020,Zhao2020a,Chu2021,Li2020,Sete2021,Stehlik2021,Collodo2020,Xu2020zz}.
By engineering the parameters, the two interaction channels cancel each other.
The interaction can either be turned on through a tunable coupler or through the cross resonant control scheme.
The second approach is to choose a hybrid qubit system with opposite anharmonicity, which allows parameter engineering to suppress the ZZ interaction.
One implementation is using a transmon and a capacitively shunt flux qubit (CSFQ)~\cite{Zhao2020,Ku2020}.
Other methods include using additional off-resonant drive \cite{Wei2021,Mitchell2021,Xiong2021} and different types of qubits have also been proposed~\cite{Finck2021}.

Most of the above works are based on the strong dispersive regime, where the resonator is only weakly coupled with the qubits.
In this regime, the ZZ interaction strength $\zeta$ is only determined by the effective interaction with the two non-computational qubit state, $\ket{20}$ and $\ket{02}$~\cite{Magesan2020}
\begin{equation}
    \label{eq:strong dispersive zeta}
    \zeta_{\textnormal{disp}} = -\frac{2J^2_{20,11}}{\Delta_1-\Delta_2+\alpha_1} + \frac{2J^2_{02,11}}{\Delta_1-\Delta_2-\alpha_2}
\end{equation}
where $\Delta_q=\omega_{q} - \omega_{c}$ is the qubit-resonator frequency detuning, $\alpha_q$ the anharmonicity and $J_{jk,j'k'}$ the effective coupling strength between the physical qubit state $\ket{jk}$ and $\ket{j'k'}$.
They are obtained by performing a leading order SWT and effectively decouple the resonator from the two qubits.
In this regime, it is impossible to achieve zero ZZ interaction unless the two anharmonicity $\alpha_q$ adopt different signs.

\begin{figure*}
    \subfloat[\label{fig:ZZ landscape}]{
    \includegraphics[width=0.65\linewidth]{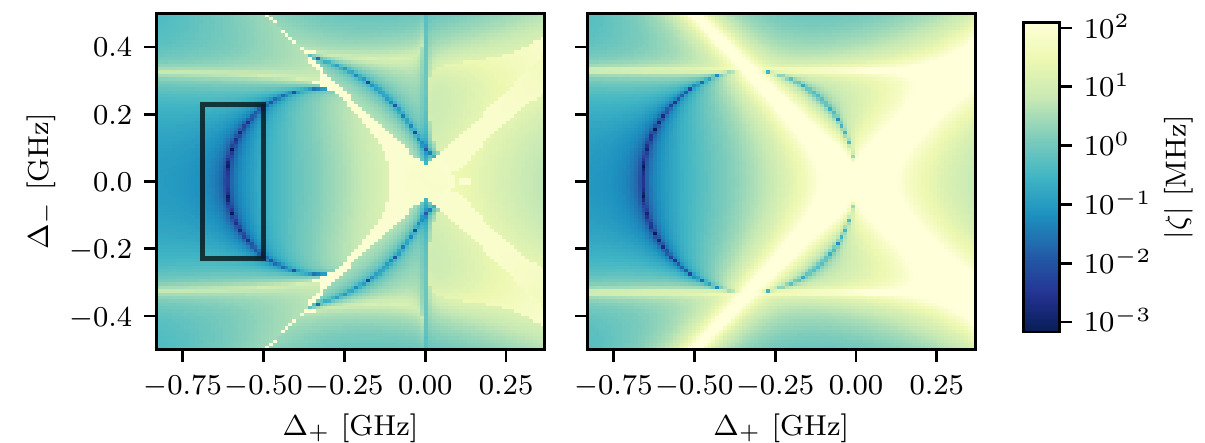}
    }
    \subfloat[\label{fig:perturbation level accuracy}]{
        \includegraphics[width=0.32\linewidth]{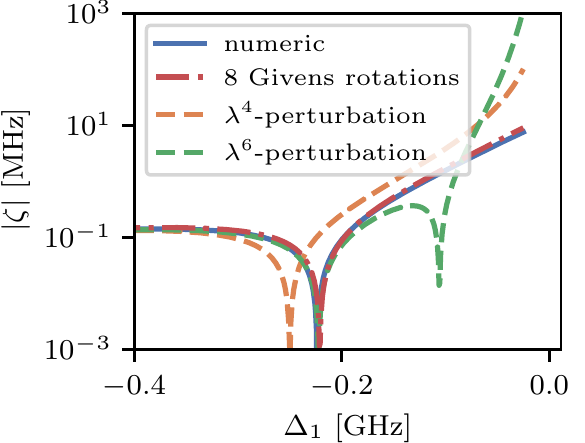}
        }
    \caption{
        {\bf(a)}: The landscape of the ZZ interaction strength $|\zeta|$ as a function of $\Delta_+=\Delta_1+\Delta_2$ and $\Delta_-=\Delta_1-\Delta_2$. {\bf Left}: numerical diagonalization of the so-called \textsc{quasidisq} regime \cite{Goerz2017}; {\bf Right}: The $\lambda^4$-perturbative approximation. 
        In the perturbative approximation, the zero points are described by a circle with a diameter of $2|\alpha|$.
        The particularly interesting regime is the left part of the circle and away from the resonant line, where the perturbation theory can still be applied, which is marked by the grey rectangle.
        In the numerical result, the circle is distorted due to the resonant lines and the left half of the circle shrinks because of the higher-order perturbative correction.
        {\bf(b)}: The numerical result compared to the perturbative correction up to $\lambda^6=(g/\Delta)^6$  and the Jacobi iteration with 8 two-by-two Givens rotations.
        Parameters used: $g_1=g_2=0.05$ GHz, $\alpha_1=\alpha_2=\alpha=-0.33$ GHz and $\Delta_{-}=0.4|\alpha|$.
    }
    \label{fig:quasi disq zeta}
\end{figure*}

However, \cref{eq:strong dispersive zeta} is only valid when ignoring the higher level of the resonator.
If we reduce the qubit detuning $\Delta_q$ so that it becomes comparable with the anharmonicity $\alpha_q$, the second excited state of the resonator comes into the picture and can be used to suppress the ZZ interaction, also known as the \textsc{quasidisq} regime~\cite{Goerz2017,Jin2021}.
We identify this regime as the quasi-dispersive regime because $g/\Delta_q$ is manufactured larger than $0.1$, e.g.~in superconducting qubits with weak anharmonicity such as Transmons, though we show the same analysis can also hold for stronger anharmonicities.
As a result, the calculation of $\zeta_{\textnormal{disp}}$ cannot be treated by only the leading-order SWT.
In particular, we will see that, in the straddling regime, where $|\Delta_1-\Delta_2| <\alpha$, the interaction with the second excited resonator state leads to a $\lambda^4$-perturbative correction that can be used to suppress the ZZ interaction.

In the following, we first use the $\lambda^4$-perturbation to qualitatively understand the energy landscape and then investigate the higher-order corrections.
For the $\lambda^4$-perturbation, using RWST, we only need 2 iterations and evaluate 4 commutators instead of 3 iterations and 11 commutators, as for traditional perturbation (\cref{tab:number of commutators}).

In fact, the traditional approach that first approximates the system as an effective qubit-qubit direct interaction and then applies another perturbation to obtain the ZZ strength is also a two-step recursion~\cite{Magesan2020}.
However, for simplicity, it neglects the resonator states in the second perturbation.
As detailed in \cref{sec:RSWT quasi-disq detail result}, adding the resonator states, we obtain a better estimation for the quasi-dispersive regime.
The result is consistent with the diagrammatic techniques used in \cite{Zhu2013,Zhao2020}.

To illustrate the energy landscape, we write the interaction strength as
\begin{equation}
    \zeta^{(4)} = g_1^2 g_2^2 \left(
    \frac{1}{\Delta_1^2(\Delta_{-} - \alpha_2)} -
    \frac{1}{\Delta_2^2(\Delta_{-} + \alpha_1)} +
    \frac{\Delta_1+\Delta_2}{\Delta_2^2\Delta_1^2}
    \right)
    \label{eq:quasi-disq zeta4}
\end{equation}
with $\Delta_{-} = \Delta_1 - \Delta_2$.
The first two terms coincide with \cref{eq:strong dispersive zeta} in the strong dispersive regime, up to $\mathcal{O}(\frac{g^4}{\Delta^3})$.

Assuming $\alpha=\alpha_1=\alpha_2$ and set $\zeta^{(4)} = 0$ in \cref{eq:quasi-disq zeta4}, we obtain an equation of a circle that describes the location of the zero points
\begin{equation}
\boxed{
    (\Delta_+ - \alpha)^2 + \Delta_-^2 - \alpha^2 = 0
    }
    \label{eq:zero points zeta4}
\end{equation}
where $\Delta_+ = \Delta_1+\Delta_2$ and $\Delta_- = \Delta_1-\Delta_2$.
In this $\lambda^4$-perturbation, the zero-points depend only on the anharmonicity $\alpha$ but not on the coupling strength $g_q$.
\Cref{eq:zero points zeta4} indicates that the ZZ interaction can be suppressed by varying the sum and difference of the two qubit-resonator detunings, as illustrated in \cref{fig:ZZ landscape}.
Because the perturbative approximation is only valid away from the resonant lines, the useful part of the parameter regime is the half-circle with $\Delta_+ < \alpha$, in particular, the region marked by the grey box in \cref{fig:ZZ landscape}.

\floatsetup[figure]{style=plain,subcapbesideposition=top}
\begin{figure}
    \sidesubfloat[ ]{
        \label{fig:contribution to zeta plot}
        \includegraphics[width=0.8\linewidth]{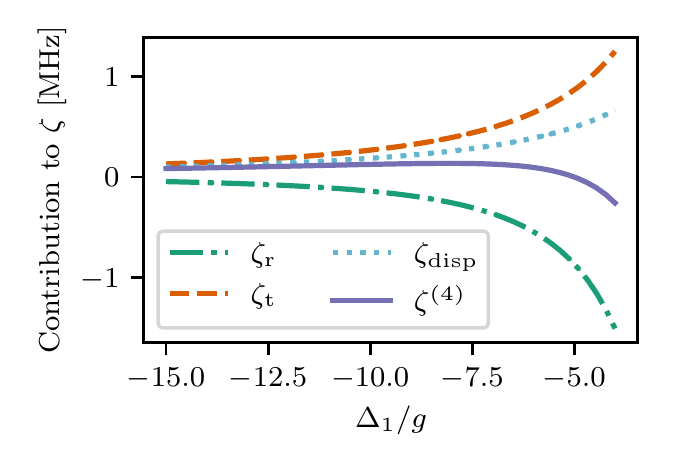}
    }\\
    \sidesubfloat[ ]{
    \label{fig:contribution to zeta illustration}
    \includegraphics[width=0.8\linewidth]{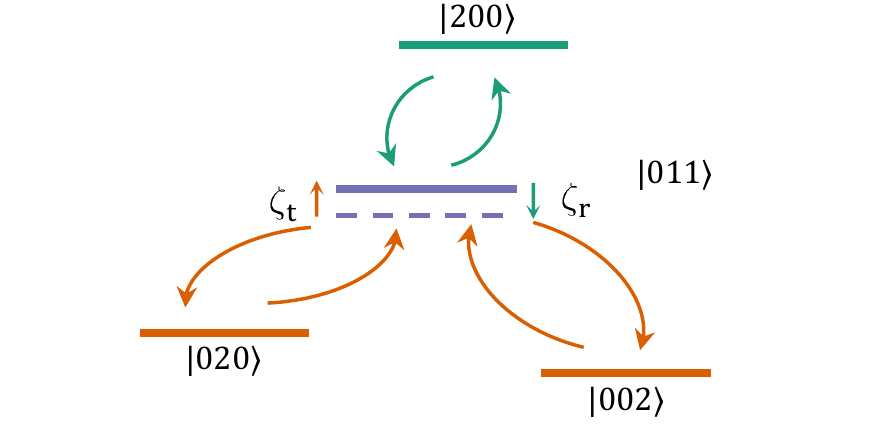}
    }
    \caption{{\bf (a)}: Different contributions to the ZZ interaction in the quasi-dispersive regime.
    The symbols $\zeta_{\rm{t}}$ and $\zeta_{\rm{r}}$ represent the contribution of virtual interaction with the second excited qubit (t) and resonator (r) states in the $\lambda^4$-perturbation.
    The former is the typical cause of ZZ cross-talk in the strong dispersive regime, while the latter is used to counteract the energy shift.
    The notation $\zeta^{(4)}$ refers to the $\lambda^4$-perturbation [\cref{eq:quasi-disq zeta4}] that goes pass zero in the quasi-dispersive regime.
    In addition, $\zeta_{\textnormal{disp}}$ denotes the strong dispersive approximation [\cref{eq:strong dispersive zeta}], which also underestimates the ZZ interaction induced by the non-qubit states. Parameters used are the same as in \cref{fig:quasi disq zeta}.
    {\bf (b)}: Illustration of the two contributions to the ZZ interaction strength in the quasi-dispersive regime.
    The solid lines and the curved arrows represent the bare states and the interaction among them.
    The second excited resonator and transmon states push the qubit $\ket{11}$ state into different directions.
    }
    \label{fig:contribution to zeta}
\end{figure}

In addition, we also studied different contributions to the ZZ interaction.
In \cref{fig:contribution to zeta plot}, we plot the strong dispersive approximation, the $\lambda^4$-perturbation as well as the contribution of second excited qubit and resonator state to $\zeta^{(4)}$ (see \cref{sec:RSWT quasi-disq detail result} for analytical expressions).
One observes from the plot that, in the quasi dispersive regime, the increasing virtual interaction with the second excited resonator state acts against the interaction with the second excited qubit states.
Notice that all contributions to $\zeta^{(4)}$ are virtual interactions of the second excited state, i.e., $\zeta^{(4)} = \zeta_{\textnormal{t}} + \zeta_{\textnormal{r}}$, as illustrated in \cref{fig:contribution to zeta illustration}.

Although the $\lambda^4$-perturbation gives insight into the different contributions to the energy shift, perturbation beyond the order $\lambda^4$ also has a non-negligible contribution in the quasi-dispersive regime.
Since RSWT requires considerably fewer commutators, we are able to compute the $\lambda^6$-perturbation, with only two iterations and 7 commutators (see \cref{tab:number of commutators}).
The $\lambda^6$-perturbation captures the location of the minimum more accurately, but still shows a false minimum close to the resonant regime, as shown in \cref{fig:perturbation level accuracy}.

Apart from perturbation, we also apply NPAD to compute the interaction strength.
We first define 4 Givens rotations with respect to the direct qubit-resonator coupling terms from the original Hamiltonian.
The rotations are then applied sequentially to obtain the first effective Hamiltonian.
Next, we apply another 4 rotations targeted at the two-photon couplings, such as the effective qubit-qubit coupling.
The indices of those 8 rotations are listed in the first two columns of \cref{tab:rotation targets}.
These two steps are equivalent to the two iterations in RSWT.
However, the recursive Givens rotations replace the BCH expansion, resulting in a much simpler calculation.
Illustrated in \cref{fig:perturbation level accuracy}, the approximation with those 8 rotations is as good as the $\lambda^6$-perturbation, but without the false minimum.
Both capture the zero points very well compared to the numerical diagonalization, where the 4 lowest levels are included for each qubit and the resonator.

With those calculations, we can then investigate the effect of the high order corrections.
We find that, for instance, $g_q$ shifts the zero point to the regime of smaller frequency detuning, corresponding to shrinking the half-circle in the numerical calculation in \cref{fig:ZZ landscape}.
In addition, for stronger coupling strength, the dip becomes narrower, which indicates a trade-off between the interaction strength and feasibility of qubit fabrication~\cite{Hertzberg2020}.
A detailed description of the effect of higher-order perturbation in the quasi-dispersive regime is presented in \cref{sec:quasi disq dependency on g and alpha}.

Overall, our investigation reveals different contributions to the ZZ interaction and provides tools to study the energy landscape in this quasi-dispersive regime.
Because of the comparably smaller detuning, operations on this regime provide stronger interactions for entangling gates, and hence may achieve a better quantum speed limit for universal gate sets, i.e.~without sacrificing local gates~\cite{Goerz2017}.

{
\setlength\heavyrulewidth{0.35ex}
\setlength{\tabcolsep}{12pt}
\begin{table}
\begin{tabular}{ cc}

\toprule
\multicolumn{2}{c}{static $H$}\\
Step 1 & Step2  \\
\midrule
010-100 & 011-200\\
001-100 & 001-010\\
011-101 & 011-002\\
011-110 & 011-020\\
\bottomrule
\end{tabular}
\quad
\begin{tabular}{ c }
\toprule

\multirow{2}{*}{driving $H_\mathrm{d}$}\\
\\
\midrule
00-10 \\
01-11 \\
10-20 \\
11-21 \\

\bottomrule
\end{tabular}
\caption{
Leading coupling terms in (block-) diagonalizing the static and the driving Hamiltonians of the cross-resonance gate, upon which the Jacobi iteration is constructed.
For the static Hamiltonian (\cref{sec:application2}), the three numbers refer to the state of the resonator, qubit 1 and qubit 2, respectively. E.g. 010-001 denotes the effective coupling between the two qubits.
For the driving Hamiltonian (\cref{sec:cross resonance}), we use the effective qubit-qubit model.
Hence only the qubit states are listed.
}
\label{tab:rotation targets}
\end{table}
}

\subsection{The cross-resonance coupling strength}
\label{sec:cross resonance}
Following the previous examples, we here study superconducting qubits under an external cross-resonance drive.
The cross-resonance interaction is activated by driving the control qubit with the frequency of the target qubit, which has been studied intensively and demonstrated in various experiments~\cite{Paraoanu2006,Rigetti2010,Chow2011,Sheldon2016,kirchhoff2018,kandala2021}.
In the two-qubit subspace, the dominant Hamiltonian term is written as a Pauli matrix $ZX$, which generates a CNOT gate up to single-qubit corrections.
Therefore, ideally, only the population of the target qubit will change after the gate operation.
The effective model is usually derived by block diagonalizing the non-qubit leakage levels as well as the population flip of the control qubit~\cite{Magesan2020,Malekakhlagh2020,Malekakhlagh2021a}.
The coupling strength is then characterized by the coefficient of the $ZX$ Hamiltonian term.

The analytical block diagonalization of the Hamiltonian is only possible when neglecting all the non-qubit levels.
Hence, perturbative expansion is often used, where the small parameter is defined as $\Omega/\Delta_{-}$, i.e., the ratio between the drive amplitude and the qubit-qubit detuning.
However, to achieve fast gates, the qubit-qubit detuning is often designed to be small, ranging from 50~MHz to 200~MHz.
Therefore, the perturbative diagonalization only works well for a weak drive.

In the rest of this subsection, we show that with only 4 two-by-two Givens rotations on the single-photon couplings, we can block-diagonalize the drive term and obtain an estimation of the coupling strength as good as the numerical result and far above the perturbative regime.

We start from the static Hamiltonian $H$ in \cref{eq:sc duffing model} and define a driving Hamiltonian in the rotating frame
\begin{equation}
    H_{\mathrm{d}} = \frac{\Omega}{2}(b_1 + b_1^{\dagger})
    .
\end{equation}
The full Hamiltonian is then written as $H+H_{\mathrm{d}}-H_{\mathrm{R}}$
where $H_{\mathrm{R}}=\omega_{d} (b_1^{\dagger} b_1 + b_2^{\dagger} b_2)$ with $\omega_{d}$ the driving frequency~\cite{Magesan2020}.
To compute the interaction strength, both the qubit-qubit effective interaction $g$ and the drive on the control qubit $\Omega$ need to be diagonalized.
In particular, the second one can be as large as the energy gap and dominant in the unwanted couplings~\cite{Malekakhlagh2021a}.
For simplicity, we assume $g$ is small and diagonalize it with a leading-order perturbation, discarding all terms smaller than $\mathcal{O}(g^2)$.
In this frame, one obtains a $ZX$ interaction that increases linearly with the drive strength~\cite{Magesan2020}.
This is equivalent to moving to the eigenbases of the idling qubits and allows us to focus on applying NPAD to the drive $H_{\mathrm{d}}$.
The same method used in \cref{sec:application2} can be applied here to improve this approximation.

Targeting the dominant drive terms listed in the right column of \cref{tab:rotation targets}, we construct 4 Givens rotations. 
The rotations are constructed with respect to the same Hamiltonians and then applied iteratively as separate unitaries.
The obtained $ZX$ interaction strength reads
\begin{align}
    \label{eq:analytical cr}
    \omega_{ZX} = & 
        g \Omega \left(
        \frac{s_1^2 c_2^2 - c_1^2}{2 \Delta_-}
        - \frac{s_2^2}{(\Delta_- + a_1)} \right.
        \\\nonumber
        + & 
        \left.
        \frac{(s_1^2 - c_1^2 c_2^2) ( a_1-\Delta_-) - \sqrt{2} a_1 s_1 s_2 c_2}{2 \Delta_- (\Delta_- + a_1)}
        \right)
\end{align}
with $c_j = \cos(\theta_j/2)$, $s_j = \sin(\theta_j/2)$ and $\Delta_- = \omega_1-\omega_2$.
The rotation angles are defined by the drive strength $\theta_1 = \arctan(\frac{\Omega}{\Delta_-})$
and
$\theta_2 = \arctan(\frac{\sqrt{2}\Omega}{2\Delta_-+a_1})$.

\begin{figure}
    \includegraphics[width=\linewidth]{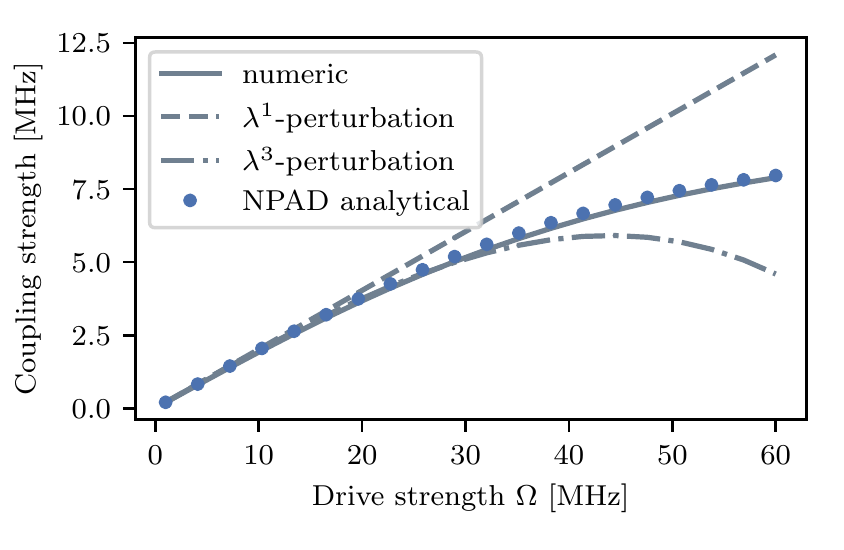}
    \caption{
        Cross-resonance coupling strength as a function of the drive strength.
        The analytical coupling strength is computed with 4 two-by-two Givens rotations on the single-photon coupling terms [\cref{eq:analytical cr}] and compared to perturbative expansion and the numerical calculation. The parameters used are inspired by the device in Ref.~\cite{kandala2021}, with the qubit-qubit detuning approximately 60~MHz and an effective qubit-qubit coupling $-3$~MHz.
        }
    \label{fig:zx coupling strength}
\end{figure}

This analytical coupling strength is plotted in \cref{fig:zx coupling strength}, compared with the perturbative expansions in Ref.~\cite{Magesan2020} and numerical block-diagonalization.
The result matches well with the numerical calculation, even when the ratio $\frac{\Omega}{\Delta_-}$ is approaching one.
On the contrary, the perturbative expansion shows a large deviation as the driving power increases.
The numerical block-diagonalization is implemented using the least action method~\cite{Cederbaum1989,Magesan2020,Xu2020zz}.
To our surprise, although no least action condition is imposed on the Jacobi iteration, the method automatically follows this track and avoids unnecessary rotations.
This suggests that the Jacobi iteration chooses an efficient path of block-diagonalization.

Notice that in the above example, no rotations are performed for levels beyond the second excited state because they are not directly coupled to the qubit subspace.
In other parameter regimes, more couplings terms may become significant and need to be added to the diagonalization.
For instance, the two-photon interaction between $\ket{0}$ and $\ket{2}$ of the control qubit will be dominant in the regime where $\Delta_- \approx -\alpha_2/2$~\cite{Malekakhlagh2020}.
The fact that high precision can be achieved with only rotations on the single-photon couplings in this example also indicates that the dominant error of perturbation lies in the BCH expansion used in diagonalizing the strong single-photon couplings, rather than in higher levels or high-order interactions.

\section{Conclusion and outlook}
\label{sec:conclusion}
We introduced the symbolic algorithm NPAD, based on the Jacobi iteration, for computing closed-form, parametric expressions of effective Hamiltonians.
The method applies rotation unitaries iteratively on to a Hamiltonian, with each rotation recursively defined upon the previous result and removing a chosen coupling between two states.
Compared to perturbation, it uses two-by-two rotations to avoid the exponentially increasing commutators in the BCH expansion.
In the perturbative limit, the method reduces to a modified form of the Schrieffer-Wolff transformation, RSWT, that inherits the recursive structure of the Jacobi iteration.
The recursive structure avoids unnecessary expansion and results in an exponential reduction in the number of commutators compared to the traditional perturbative expansion.
The two methods can also be combined as a hybrid method, where NPAD is used to remove strong couplings while RSWT is applied afterwards to effectively eliminate the remaining weak coupling.

Applying these methods to superconducting qubit systems, we showed that high precision estimation can be achieved beyond the perturbation regime, either as explicit short analytical expressions, or closed-form parametric expressions for computer-aided calculation.
Although in the study we used the Kerr model, more detailed models such as in Ref.~\cite{Malekakhlagh2020} can also be incorporated with little additional effort.

Despite the fact that using the Jacobi iteration for machine-precision diagonalization is less efficient than other methods such as QR diagonalization, the iteration can be truncated for symbolic approximation.
For many questions in quantum engineering, the most part of the energy structure and dominant couplings is known in advance.
Therefore, the iterative method can be designed for removing dominant couplings and decoupling a subspace from non-relevant Hilbert spaces, which is often used in modeling dynamics in large quantum systems~\cite{Baker2018, Gualdi2013}.
The result is, however, always a closed-form, parametric expression,
which, though usually harder for the human to read, shows its own advantage in computer-aided calculations.

We expect our method to have significant application in quantum technologies, where elimination of auxiliary or unwanted spaces (e.g. for block-diagonalization) needs to be done to significant precision to enable practically useful models. In particular, relevant applications include experiment and architecture design, reservoir engineering, cross-talk suppression, few- and many-body interaction engineering, effective qubit models, and more generally improved approximations where Schrieffer-Wolff methods are typically used. We also expect that the methods presented here will find extensions for simplifying other equations of motion, such as in open-quantum systems \cite{de2017dynamics,schirmer2010stabilizing}, non-linear systems \cite{antoine2013computational}, or for uncertainty propagation \cite{dalgaard2022dynamical}. Last but not least, accurate, parametric diagonalization should be especially useful for time-dependent diagonalization where adiabatic following can be enforced by DRAG \cite{Theis2018, Motzoi2009} or other counter-diabatic \cite{Guery2019, unanyan1997laser} approaches.

\begin{acknowledgments}
This work was funded by the Federal Ministry of Education and
Research (BMBF) within the framework programme "Quantum technologies – from
basic research to market" (Project QSolid, Grant No.~13N16149),
by the Deutsche Forschungsgemeinschaft (DFG, German Research Foundation) under Germany’s
Excellence Strategy – Cluster of Excellence Matter and
Light for Quantum Computing (ML4Q) EXC 2004/1 –
390534769, and through the European Union’s Horizon
2020 research and innovation programme under Grant
Agreements No.~817482 (PASQuanS) and No.~820394
(ASTERIQS).
\end{acknowledgments}

\appendix
\section{The error bound for truncating the BCH expansion}
\label{sec:RSWT error bound}
In the main text, we presented \cref{eq:RSWT equation} as the expression to compute the transformed matrix $H'$, which is a function of the off-diagonal part of the original matrix $V$ and the generator $S$.
The expression is derived from a truncated BCH formula.
In the following, we derive the error bound of the truncation.

Without truncation, \cref{eq:RSWT equation} is written as
\begin{equation}
    H'_{\rm{ideal}} = D + \sum_{t=1}^{\infty} \frac{t}{(t+1)!} \mathcal{C}_t(S, V)
\end{equation}
where we neglected the index $n$ for the iteration step.
If the expansion is truncated at $t=m-1$, one obtains
\begin{align}
    \epsilon = &
    \norm{
        H'_{\rm{ideal}} - H'_{\rm{trunc}}
    }
    =
    \norm{
        \sum_{t=m}^{\infty} \frac{t}{(t+1)!} \mathcal{C}_t(S, V)
    }
    \\ \nonumber
    \le &
    \sum_{t=m}^{\infty}
    \frac{t2^t}{(t+1)!}
    \norm{S}^t \norm{V}
    \le
    \frac{2^m}{m!}
    \frac{\norm{S}^m}{1-\norm{S}}
    \norm{V}
\end{align}
where we assume in the last inequality that $\norm{S}<1/2$.

\section{Efficiency comparison between RSWT and SWT}
\label{sec:number of commutators}
We show here that, given a finite-dimensional Hamiltonian $H$, RSWT is more efficient than SWT for perturbation beyond level $\lambda^2$ with an exponential decrease in the number of commutators.
We measure the complexity by the number of commutators that need to be evaluated to compute all eigenenergy corrections up to $\lambda^K$, denoted by $\mathcal{N}$.
The general formula is presented below while the numbers for $K \le 8$ is given in \cref{tab:number of commutators} in the main text.

For SWT, one can find the general expression as well as explicit formulas up to $\lambda^5$ in Ref.~\cite{Magesan2020}.
The number of iterations required to reach order $\lambda^K$ is $K-1$.
In addition, at each iteration $n$, one needs to include also mixed terms composed of generator $S_l$ with $l \le n$.
The number $\mathcal{N}$ is given by
\begin{equation}
    \mathcal{N}_{\rm{SWT}}
    =
    \sum_{n=1}^{K-1} \sum_{l=1}^n 2^{l-1}
    =
    2^K-K-1
\end{equation}
where $2^{l-1}$ is the number of distinct tuples $(S_{i_1}, S_{i_2}, S_{i_3}, \cdots)$ with $\sum_j{i_j}=l$.
We have taken into consideration that $[S_1, \rm{diag}(H)]=-V$ and $[S_{n+1}, \rm{diag}(H)]$ is known by the construction of $S_{n+1}$.

For RSWT, the calculation of commutators in each iteration is given in \cref{eq:RSWT equation}.
Because $\mathcal{C}_{t+1}(A,B)$ can be calculated from $\mathcal{C}_{t+1}(A,B)$ with only one additional commutators, the number commutators to be evaluated in \cref{eq:RSWT equation} is exactly $m-1=\lfloor \frac{K}{2^n}\rfloor-1$.
The total number of iteration $n_{\rm{max}}$ is given by $\lfloor \log_2(K)\rfloor$.
Therefore, we obtain
\begin{equation}
    \mathcal{N}_{\rm{RSWT}} = \sum_{n=0}^{\lfloor \log_2(K)\rfloor-1} \lfloor \frac{K}{2^n}\rfloor - 1 < 2K
    .
\end{equation}
The reduction compared to SWT comes from the fact that the energy difference in $H_n$ is used in the definition of $S_{n+1}$, rather than the bare energy difference in $H$. The recursive expressions avoid unnecessary expansions.
One obtains the same final expressions as from SWT up to $\lambda^K$,
if one expands the energy difference into a polynomial series
\begin{equation}
    \frac{1}{
        \Delta E_{\rm{bare}} + \Delta E_{\rm{correction}}
    }
    = 
    \frac{1}{\Delta E_{\rm{bare}}}
    \rm{poly}
    \left(
        \frac{
            \Delta E_{\rm{correction}}
        }{
            \Delta E_{\rm{bare}}
        }
    \right)
\end{equation}
and substitutes in expressions so that it depends only on the bare energy and couplings.

\begin{figure}
	\centering
    \includegraphics{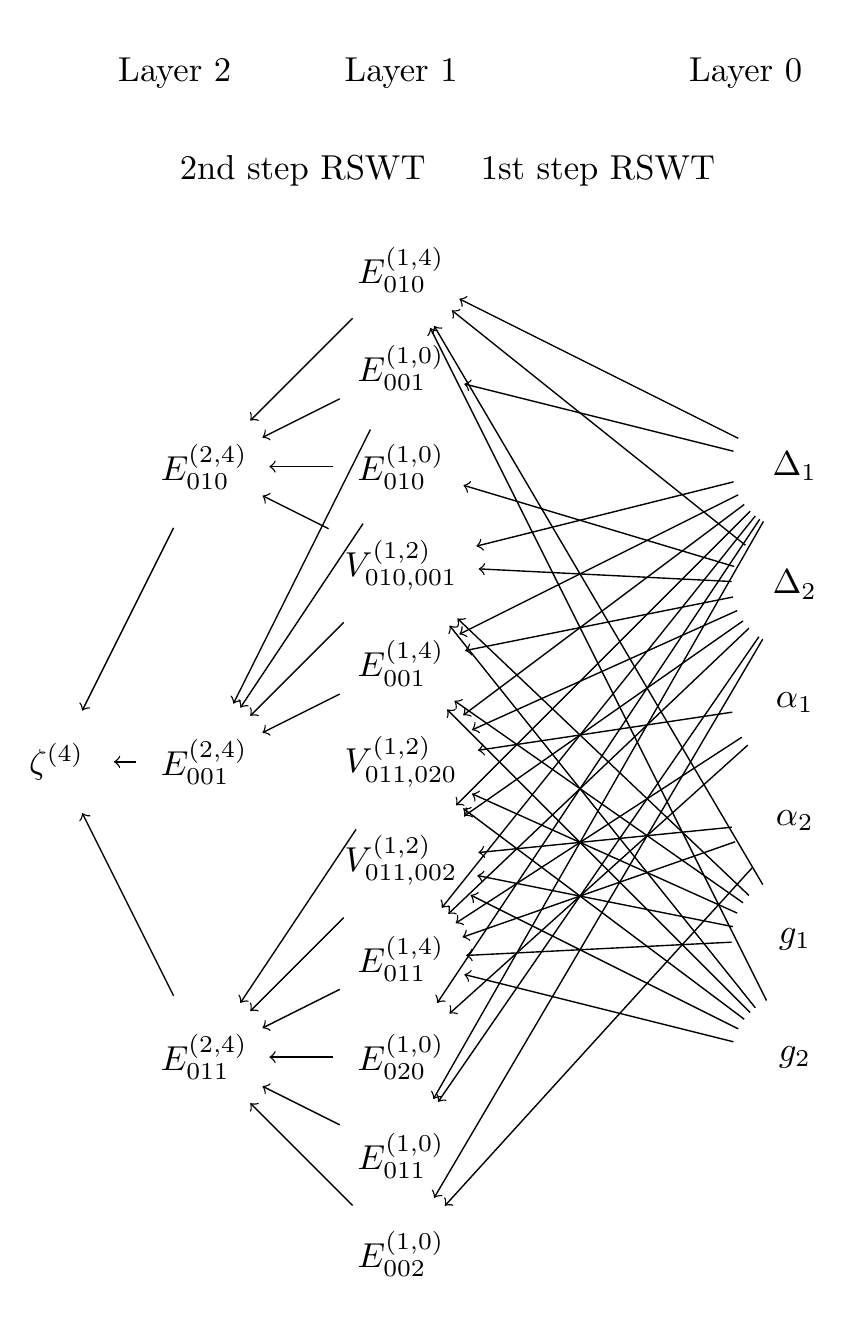}
	\caption{The network illustration of the clossed-form expression of $\zeta^{(4)}$ parameterized by $\Delta_1$, $\Delta_2$, $\alpha_1$, $\alpha_2$, $g_1$, and $g_2$, obtained from the two-step RSWT.
    Each node in the 1st and 2nd layers is a matrix entry in the Hamiltonian $H_1$ and $H_2$.
    A node in layer $n+1$ is expressed as a function of the nodes in layer $n$, represented by an edge. 
    In particular, symbols $E^{(n,k)}_{lpq}$ represent the $\lambda^k$ diagonal entries of $\bra{lpq} H_n\ket{lpq}$ and $V_{lpq,l'p'q'}^{(n,k)}$ the effective coupling.
    The upper index $k$ denotes the level of perturbation, e.g., $k=4$ means that it is a $\lambda^4$-perturbative correction.
    }
	\label{fig:recursive tree for zeta4}
\end{figure}

\section{RSWT results for the ZZ interaction strength}
\label{sec:RSWT quasi-disq detail result}

Using RSWT described in \cref{sec:RSWT}, we compute the effective Hamiltonian up to $\lambda^6$, where $\lambda$ is defined as the ratio between the largest coupling and energy gap.
To compute the $\lambda^4$- and $\lambda^6$-perturbation, RWST only takes 2 iterations with 4 and 7 commutators respectively, which is significantly smaller than those required for SWT as shown in \cref{tab:number of commutators}.
A third iteration only adds an improvement of $\mathcal{O}(\lambda^8)$ to the eigenenergy because the off-diagonal terms of $H_2$ are at most $\mathcal{O}(\lambda^4)$.

Because of the recursive structure of RSWT, each matrix element in $H_{n+1}$ is given as a function of matrix elements in $H_n$.
Hence, the final result is a closed-form expression parameterized by the matrix elements of the original Hamiltonian $H$, i.e. the hardware parameters.
The parametric expression consists only of algebraic expressions and the dependence can be illustrated as a network.
For instance, we show the network representation of the $\lambda^4$-perturbation $\zeta^{(4)}$ in \cref{fig:recursive tree for zeta4}.
Each symbol in layer $n+1$ is analytically expressed as a function of symbols in layer $n$, represented by arrows.
The arrows between the first and the second layer represent the definition $\zeta^{(4)}=E^{(4)}_{011}-E^{(4)}_{001}-E^{(4)}_{010}$.
Given all the six hardware parameters (layer 0), one can evaluate $\zeta^{(4)}$ by recursively evaluating all the nodes it depends on.

In the following, we present the analysis of $\lambda^4$- and $\lambda^6$-perturbation.

\subsection{$\lambda^4$-perturbation}
The $\lambda^4$-perturbative correction for $\zeta$ is given as
\begin{equation}
    \zeta^{(4)} = E_{011}^{(2,4)} - E_{010}^{(2,4)} - E_{001}^{(2,4)}
    .
\end{equation}
The notation $E_{lpq}^{(n,k)}$ represents the $\lambda^k$-perturbation obtained from $H_n$.
The sub-indices $lpq$ denotes the resonator state $\ket{l}$ and two qubit states $\ket{p}$, $\ket{q}$.

We first calculate $E_{011}^{(2,4)}$. Substituting the expression for $H_2$ as a function of entries in $H_1$, we obtain
\begin{align}
\label{eq:RSWT map H1 to H2}
    E_{011}^{(2,4)} =  &
    \frac{V_{002,011}^{(1,2)} V_{011,002}^{(1,2)}}{ E_{011}^{(1,0)} - E_{002}^{(1,0)}}
    +\frac{V_{011,020}^{(1,2)} V_{020,011}^{(1,2)}}{ E_{011}^{(1,0)} - E_{020}^{(1,0)}}
    \\ \nonumber&
    +\frac{V_{011,200}^{(1,2)} V_{200,011}^{(1,2)}}{ E_{011}^{(1,0)} - E_{200}^{(1,0)}}
    + E_{011}^{(1,4)}
\end{align}
where $V_{lpq,l'p'q'}^{(n,k)}$ denotes the interaction between state $\ket{lpq}$ and $\ket{l'p'q'}$.

The physical meaning of each term in \cref{eq:RSWT map H1 to H2} can be interpreted as follows:
The first two terms are identical to the dispersive approximation given in \cref{eq:strong dispersive zeta}, which is the consequence of the effective qubit-qubit interaction.
The third term, depending on the effective interaction between $\ket{200}$ and $\ket{011}$, is 0 at this order.
This is because the destructive interference between the path $\ket{011} \rightarrow \ket{110} \rightarrow \ket{200}$ and $\ket{011} \rightarrow \ket{101} \rightarrow \ket{200}$ results in $V_{011,200}^{(1,2)}=V_{200,011}^{(1,2)}=0$.
The last term, $E_{011}^{(1,4)}$, is what the approximation of a strong dispersive regime fails to characterize.
It was generated by the commutator $[S_1, [S_1, [S_1, V_0]]]$ and the energy gaps in the denominator of entries in $S_1$ are always the qubit-resonator detuning (plus the anharmonicity), which, in the strong dispersive regime, is much larger than the qubit-qubit detuning in \cref{eq:strong dispersive zeta}.
Hence the last term is much smaller in the strong dispersive regime.
However, in the quasi-dispersive regime, it plays a key role in suppressing the ZZ interaction as shown in \cref{fig:contribution to zeta illustration}.

After including the single-excitation terms $E_{010}^{(2,4)}$ and $E_{001}^{(2,4)}$ using the same two-step RSWT, we separate the contributions of virtual interaction into 2 categories: those including the second excited qubit state (denoted by t) and those including the second excited resonator state (denoted by r):
\begin{align}
    \zeta^{(4)}_{\textnormal{t}}
    = \nonumber &
    \zeta_{\textnormal{disp}}
    - \frac{g_{1}^{2} g_{2}^{2}}{2 \Delta_{2}  \left(\Delta_{1} + \alpha_{1}\right)^2}
    - \frac{3 g_{1}^{2} g_{2}^{2}}{2 \Delta_{2}^{2} \left(\Delta_{1} + \alpha_{1}\right)}
    \\ &
    - \frac{g_{1}^{2} g_{2}^{2}}{2 \Delta_{1}  \left(\Delta_{2} + \alpha_{2}\right)^2}
    - \frac{3 g_{1}^{2} g_{2}^{2}}{2 \Delta_{1}^{2} \left(\Delta_{2} + \alpha_{2}\right)}
    \\
    \zeta^{(4)}_{\textnormal{r}}
    = &
    \frac{2 g_{1}^{2} g_{2}^{2}}{\Delta_{1} \Delta_{2}^{2}} + \frac{2 g_{1}^{2} g_{2}^{2}}{\Delta_{1}^{2} \Delta_{2}}
\end{align}
where $\zeta_{\textnormal{disp}}$ is given by \cref{eq:strong dispersive zeta}.
Summing all the contributions gives the $\lambda^4$-perturbation $\zeta^{(4)}$ in \cref{eq:quasi-disq zeta4}.
Notice that virtual interactions that only involve the first excited state have no contribution to the ZZ interaction at this perturbation level, i.e., $\zeta^{(4)} = \zeta_{\textnormal{t}} + \zeta_{\textnormal{t}}$.
This is because the energy shift of $\ket{011}$ induced by $\ket{101}$ and $\ket{110}$ cancels that of $\ket{010}$ and $\ket{001}$ induced by $\ket{100}$.

\subsection{$\lambda^6$-perturbation}

Using the two-step RSWT, we also computed the $\lambda^6$-perturbative correction to the ZZ interaction strength:
\begin{equation}
    \zeta^{(6)} = \zeta^{(6)}_{\textnormal{disp}} + \zeta^{(6)}_{\textnormal{rest}}
\end{equation}

The first contribution corresponds to the effective qubit-qubit interaction and dominants in the strong dispersive regime.
It turns out that it only includes the next order of effective interaction and energy difference.
Hence, for simplicity, we present it together with $\zeta^{(4)}_{\textnormal{disp}}$:

\begin{widetext}
\begin{align}
    \zeta^{(4)}_{\textnormal{disp}} + \zeta^{(6)}_{\textnormal{disp}}
    =& \frac{\left(V_{011,020}^{(1,2)} + V_{011,020}^{(1,4)}\right) \left(V_{020,011}^{(1,2)} + V_{020,011}^{(1,4)}\right)}{\Delta E^{(2,0)}_{011,020} + \Delta E^{(2,2)}_{011,020}} + \frac{\left(V_{002,011}^{(1,2)} + V_{002,011}^{(1,4)}\right) \left(V_{011,002}^{(1,2)} + V_{011,002}^{(1,4)}\right)}{\Delta E^{(2,0)}_{011,002} + \Delta E^{(2,2)}_{011,002}}
\end{align}
with terms regarding to the virtual interactions between states $\ket{011}$ and $\ket{020}$ given by
\begin{align}
    V_{011,020}^{(1,2)} = V_{020, 011}^{(1,2)} =& 
    \frac{\sqrt{2} g_{1} g_{2}}{2 \left(\alpha_1 + \Delta_{1}\right)} + \frac{\sqrt{2} g_{1} g_{2}}{2 \Delta_{2}}
    ,\\
    V_{011,002}^{(1,4)} = V_{002, 011}^{(1,4)} = & 
    - \frac{\sqrt{2} g_{1} g_{2}^{3}}{4 \left(\alpha_2 + \Delta_{2}\right)^{3}} + \frac{\sqrt{2} g_{1} g_{2}^{3}}{8 \Delta_{2}^{2} \left(\alpha_2 + \Delta_{2}\right)} - \frac{7 \sqrt{2} g_{1} g_{2}^{3}}{4 \Delta_{1} \left(\alpha_2 + \Delta_{2}\right)^{2}}
     \nonumber\\ 
     & + \frac{3 \sqrt{2} g_{1} g_{2}^{3}}{2 \Delta_{1} \Delta_{2} \left(\alpha_2 + \Delta_{2}\right)} - \frac{5 \sqrt{2} g_{1} g_{2}^{3}}{8 \Delta_{1} \Delta_{2}^{2}} - \frac{7 \sqrt{2} g_{1}^{3} g_{2}}{8 \Delta_{1}^{2} \left(\alpha_2 + \Delta_{2}\right)} - \frac{\sqrt{2} g_{1}^{3} g_{2}}{8 \Delta_{1}^{3}}
    ,\\
    \Delta E^{(2,0)}_{011,020} + \Delta E^{(2,2)}_{011,020} =&
    - \alpha_1 - \Delta_{1} + \Delta_{2} - \frac{2 g_{1}^{2}}{\alpha_1 + \Delta_{1}} + \frac{g_{2}^{2}}{\Delta_{2}} + \frac{g_{1}^{2}}{\Delta_{1}}
    .
\end{align}

Terms corresponding to states $\ket{011}$ and $\ket{002}$ are obtained by interchanging the sub-index 1 and 2 in each expression above.

The rest of the contribution can be summed as
\begin{equation}
    \zeta^{(6)}_{\textnormal{rest}} = 
    \zeta^{(6)}_{\textnormal{rest}, g_{1}^{2} g_{2}^{4}} + \zeta^{(6)}_{\textnormal{rest}, g_{1}^{4} g_{2}^{2}}
\end{equation}
with

\begin{align}
   \zeta^{(6)}_{\textnormal{rest}, g_{1}^{2} g_{2}^{4}}
   = & \frac{9 g_{1}^{2} g_{2}^{4}}{4 \Delta_{2}^{3} \left(\Delta_{1} + \alpha_{1}\right)^{2}}
   + \frac{23 g_{1}^{2} g_{2}^{4}}{4 \Delta_{2}^{4} \left(\Delta_{1} + \alpha_{1}\right)}
   + \frac{g_{1}^{2} g_{2}^{4}}{2 \Delta_{1} \left(\Delta_{2} + \alpha_{2}\right)^{4}}
   - \frac{g_{1}^{2} g_{2}^{4}}{4 \Delta_{1} \Delta_{2}^{2} \left(\Delta_{2} + \alpha_{2}\right)^{2}}
   - \frac{4 g_{1}^{2} g_{2}^{4}}{\Delta_{1}^{3} \Delta_{2} \left(\Delta_{2} + \alpha_{2}\right)}
    \nonumber \\
   + & \frac{7 g_{1}^{2} g_{2}^{4}}{2 \Delta_{1}^{2} \left(\Delta_{2} + \alpha_{2}\right)^{3}}
   - \frac{5 g_{1}^{2} g_{2}^{4}}{2 \Delta_{1}^{2} \Delta_{2} \left(\Delta_{2} + \alpha_{2}\right)^{2}}
   + \frac{3 g_{1}^{2} g_{2}^{4}}{4 \Delta_{1}^{2} \Delta_{2}^{2} \left(\Delta_{2} + \alpha_{2}\right)}
   - \frac{4 g_{1}^{2} g_{2}^{4}}{\Delta_{1}^{2} \Delta_{2}^{3}}
   + \frac{4 g_{1}^{2} g_{2}^{4}}{\Delta_{1}^{3} \left(\Delta_{2} + \alpha_{2}\right)^{2}}
   - \frac{6 g_{1}^{2} g_{2}^{4}}{\Delta_{1} \Delta_{2}^{4}}
   .
\end{align}
\end{widetext}
The second contribution, $\zeta^{(6)}_{\textnormal{rest}, g_{1}^{4} g_{2}^{2}}$, is obtained again by interchanging the sub-index 1 and 2.

\begin{figure*}
    \centering
    \includegraphics[width=0.95\linewidth]{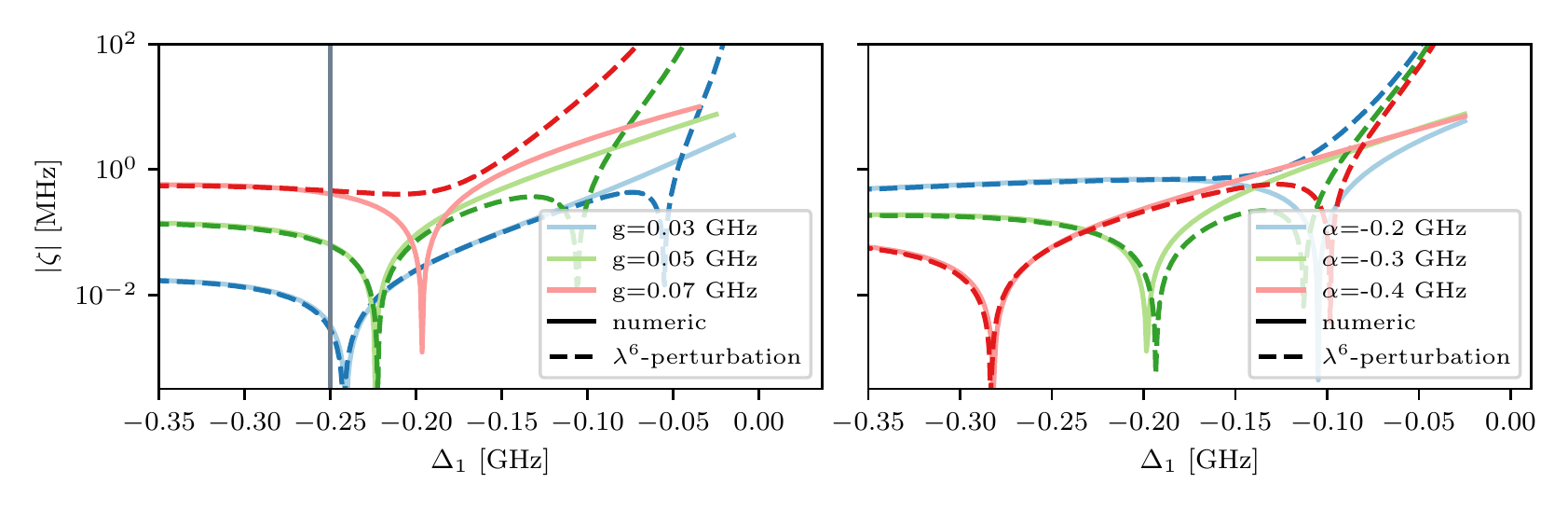}
    \caption{The dependency of $\zeta$ on the resonator-qubit interaction strength $g$ and the qubit anharmonicity $\alpha$.
    Computed with RSWT to the $\lambda^6$-perturbation.
    {\bf Left:} Dependency on $g$. The vertical line denotes the zero point predicted by the 4th-order perturbation, which is independent on $g$. Both the numerical result and the $\lambda^6$-perturbation indicate that the zero points are shifted to the regime with smaller qubit-resonator detuning.
    {\bf Right:} Dependency on $\alpha$.
    The default parameters used, if not specified in the plots, are $\Delta_-=0.4|\alpha|$, $g=50$ MHz, $\alpha_1=\alpha_2=\alpha=-330$ MHz.
    }
    \label{fig:quasi disq dependency on g and alpha}
\end{figure*}

\section{Effect of higher-order perturbation on the zero points of ZZ interaction}
\label{sec:quasi disq dependency on g and alpha}

The $\lambda^4$-perturbation described by \cref{eq:quasi-disq zeta4} predicts the zero points as a circle with a radius of $2|\alpha|$, independent of $g$.
However, as they are located in the quasi-dispersive regime for systems with weak anharmonicity, the higher-order perturbation is not always negligible.
Here, we qualitatively describe how the higher-order ($>4$) affect the zero points of $\zeta$. 

We observe that, in contrast to the $\lambda^4$-perturbation, when including the higher orders, the zero points depends on the coupling strength $g$.
As shown in \cref{fig:quasi disq dependency on g and alpha}, the higher-order perturbation shifts the zero points to the regime of smaller detuning.
The larger the coupling, the stronger the shift is.

One can estimate the accuracy of perturbation around the zero points by the ratio $g/|\alpha|$.
At the zero points of $\zeta$, the larger the anharmonicity and the smaller the coupling, the better is the perturbative approximation.
This is because the perturbation is characterized by the small parameter $\lambda=g/\Delta$ and near the zero-points $\Delta$ depends linearly on $\alpha$ [see \cref{eq:quasi-disq zeta4}],
hence the ratio $g/|\alpha|$.
This is also illustrated in \cref{fig:quasi disq dependency on g and alpha}, where we compare the deviation between the numerical result and the perturbation.
The minimum even vanishes in the analytical result when it is close to the resonant lines.
This behaviour also indicates that for superconducting qubits with a relatively large anharmonicity, the ZZ interaction can also be completely suppressed in the strong dispersive regime in this qubit-resonator-qubit model.

\end{document}